\documentclass[final, a4paper]{aastex7}

\usepackage{color}
\usepackage{enumitem}
\usepackage{hyperref}
\usepackage{verbatim}
\usepackage{amsmath,amstext,mathrsfs}
\usepackage[all]{hypcap} 
\usepackage{afterpage}    
\usepackage{marginnote}
\usepackage{makecell}
\usepackage{subfigure}
\usepackage{multirow}

\shorttitle{$B-\rho$ slope and ${\cal M}_{\rm A}$}
    \shortauthors{Zhao, Li $\&$ Qiu}
\usepackage{enumitem}
\setlist[enumerate]{listparindent=\parindent}

\begin{document}

\title{A Unified Magnetohydrodynamic Scaling Relation for the Multiphase Interstellar Medium }

\correspondingauthor{Guang-Xing Li, Keping Qiu}
\email{gxli@ynu.edu.cn,ligx.ngc7293@gmail.com, kpqiu@nju.edu.cn}

\author[0000-0003-0596-6608]{Mengke Zhao}
\affil{School of Astronomy and Space Science, Nanjing University, 163 Xianlin Avenue, Nanjing 210023, Jiangsu, People’s Republic of China}
\affil{South-Western Institute for Astronomy Research, Yunnan University, Kunming 650091, People’s Republic of China}
\affil{Key Laboratory of Modern Astronomy and Astrophysics (Nanjing University), Ministry of Education, Nanjing 210023, Jiangsu, People’s Republic of China}
\email{mkzhao@nju.edu.cn}

\author[0000-0003-3144-1952]{Guang-Xing Li}
\affil{South-Western Institute for Astronomy Research, Yunnan University, Kunming 650091, People’s Republic of China}
\email{gxli@ynu.edu.cn,ligx.ngc7293@gmail.com}

\author[0000-0002-5093-5088]{Keping Qiu}
\affil{School of Astronomy and Space Science, Nanjing University, 163 Xianlin Avenue, Nanjing 210023, Jiangsu, People’s Republic of China}
\affil{Key Laboratory of Modern Astronomy and Astrophysics (Nanjing University), Ministry of Education, Nanjing 210023, Jiangsu, People’s Republic of China}
\email{kpqiu@nju.edu.cn}

\begin{abstract}
The interplay of magnetic fields, turbulence, and gravity governs the structural evolution of the interstellar medium (ISM) and the initial conditions of star formation, yet observational gaps have long enforced a ``broken power-law'' description of the magnetic field--density relation.
Here we assemble a unified dataset spanning ten orders of magnitude in density ($10^{-26}$--$10^{-16}\,\mathrm{g\,cm^{-3}}$) by combining pulsar and Zeeman observations.
The unified data are consistent with a single, continuous magnetic-field evolution organised by the Alfv\'en Mach number $\mathcal{M}_{\rm A}=\sqrt{E_K/E_B}$. Within this interpretation, the low-density gas is magnetically dominated ($\mathcal{M}_{\rm A}<1$), whereas the high-density gas becomes kinetically dominated ($\mathcal{M}_{\rm A}>1$) as gravity increasingly contributes to the kinetic-energy budget, with magnetic tension continuing to influence the collapse geometry.
Within the Gradual Transition interpretation, the empirical break density traces the vicinity of the trans-Alfv\'enic equipartition point, $\mathcal{M}_{\rm A}=1$.
This Gradual Transition model makes three predictions tested here.
Its exponent and background field are fixed in advance by turbulent physics and recovered by the fits ($\beta\approx0.15$--$0.21$ against a predicted $0.147$; $B_c\approx2.0\,\mu$G).
A dense-gas fit, extrapolated blindly across four decades, passes through the diffuse pulsar data.
And, under the adopted scale mappings, the implied magnetic-energy spectrum approaches a $k^{-5/3}$-like scaling on large scales and departs from this extrapolation on small scales, where gravitational compression amplifies the field.
The broken power law can therefore be viewed as a piecewise approximation to the continuous magnetic equation of state, with its fitted transition density potentially retaining a physical connection to the onset of gravity-driven motions.
\end{abstract}

\section{Introduction}

\begin{figure}[h]
    \centering
    \includegraphics[width=0.95\linewidth]{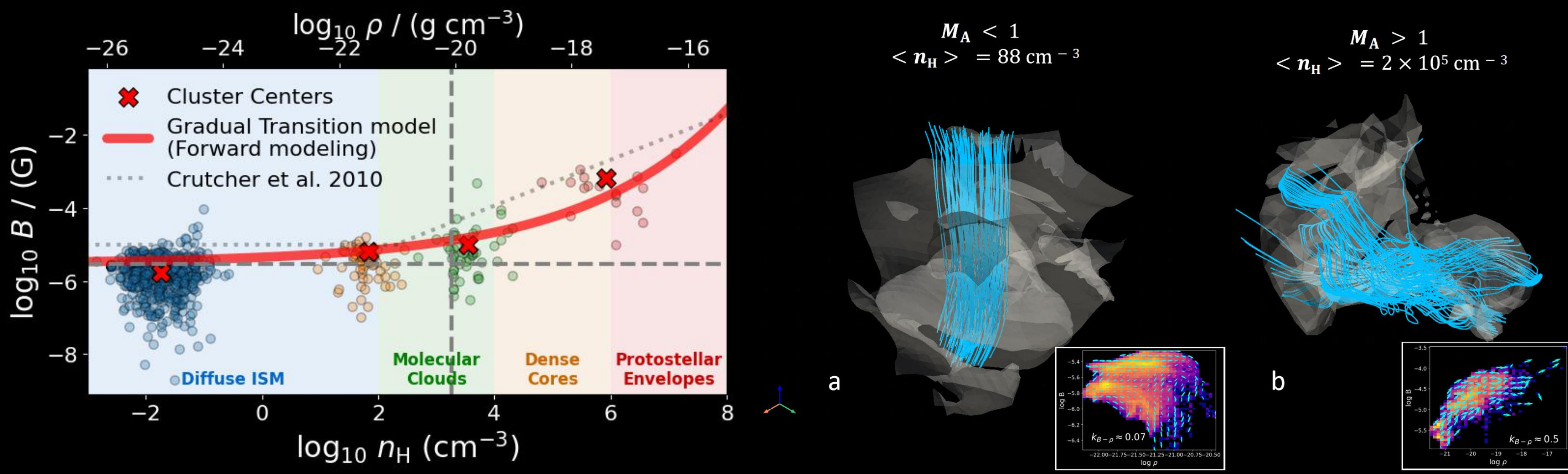}
    \caption{{\bf Gradual Transition $B-\rho$ relation and different Alfvénic behaviors.}
    }
    \label{figoverview}
\end{figure}

Deriving a continuous physical law for the interstellar magnetic field has long been obstructed by the discrete nature of observational tracers \citep{2010ApJ...725..466C}.
While the transition from turbulence-dominated diffuse gas to self-gravitating cores is theoretically expected to be continuous \citep{2012A&ARv..20...55H}, observational limitations have enforced a segmented view \citep{2012ARA&A..50...29C, 2023ASPC..534..193P}. 
Traditional Zeeman surveys rely on specific tracers (H\textsc{i}, OH, CN), resulting in a clustered sampling of the density dynamic range ($n_{\rm H} \gtrsim 10$ cm$^{-3}$) separated by significant observational gaps \citep{2010ApJ...725..466C}. 
The Bayesian framework for analyzing these measurements was established by \citet{2010ApJ...725..466C}.
However, restricted by the sensitivity limits of that era to dense molecular tracers, a 'broken power-law' emerged as the most parsimonious description of the available data.
This model separates a density-independent regime ($B \propto \rho^0$) from a collapse-driven power-law regime ($B \propto \rho^{2/3}$; \citealt{1966MNRAS.133..265M}) at an empirical break density $n_0 \approx 300$ cm$^{-3}$ \citep{2010ApJ...725..466C}. 
While providing a robust fit to the dense gas region, such piecewise descriptions impose a discontinuity that may not reflect the underlying physics \citep{2025MNRAS.tmp..512S}.
The form of this relation, however, is sensitive to how observational uncertainties are treated: relaxing the density errors lowers the high-density slope from $k_{B-\rho} \approx 2/3$ towards $k_{B-\rho} \approx 1/2$ \citep{2015MNRAS.451.4384T}, and a hierarchical Bayesian treatment of both field and density uncertainties leaves the transition density poorly constrained \citep{2020ApJ...890..153J}.

Two recent developments motivate a continuous $B(\rho)$. 
A visual overview of the proposed magnetic evolution is shown in Fig.~\ref{figoverview}.
On the theoretical side, \citet{2024ApJ...976..209Z} established in MHD simulations the fundamental relation between the $B$-$\rho$ slope and the Alfvén Mach number ($\mathcal{M}_{\rm A}$), the dimensionless ratio of kinetic to magnetic energy, from which a continuous Gradual Transition (GT) form of the $B-\rho$ relation follows. 
On the observational side, pulsar dispersion measures have been combined with Zeeman data to extend the relation into the diffuse ionised medium \citep{2025MNRAS.tmp..512S}, and a hierarchical Bayesian framework incorporating pulsars and an explicit density-uncertainty parameter has been applied within the broken power-law picture \citep{2026arXiv260306838W}.
What has not been done is to check the continuous GT form directly against such a unified dataset, and to ask whether the continuity is intrinsic to the magnetic evolution or an artefact of joining different tracers.
The novelty of the present work is therefore not the identity $\mathcal{M}_{\rm A}^2=E_K/E_B$ itself, but a parameter-level observational test of the continuous $B(\rho)$ relation derived from it. We additionally examine the contrasting magnetic-field morphology and $B-\rho$ coherence on the two sides of the trans-Alfv\'enic transition.
This is the question we address here. 
We fit the GT model to a dataset spanning ten orders of magnitude in density, infer $\mathcal{M}_{\rm A}$ from the $B-\rho$ slope through the simulation-calibrated fundamental relation, and quantify the resulting parameters with their uncertainties. 
In this work, we confront the Gradual Transition model with the data on several fronts: an externally calibrated forward prediction with no parameters fitted to the present dataset, a Bayesian model comparison against the broken power law across different sample compositions and density-error assumptions, and a blind extrapolation that predicts the diffuse-gas behaviour from dense-gas data alone. 
A model-independent local-slope analysis tests the qualitative shape of the relation at the densities where the historical $k_{B-\rho} = 2/3$ debate has been most active. 
Building on earlier magnetic-energy-spectrum formulations for molecular clouds \citep{2018MNRAS.474.2167L}, we translate the continuous B-$\rho$ relation into wavenumber space and derive the magnetic energy spectrum it predicts, and identify a systematic departure from the Galactic Big Power Law at small scales, set by the same characteristic density that governs the $B-\rho$ relation.

\section{A Continuous $B-\rho$ Relation Across Ten Orders of Magnitude}

\subsection{The Unified Dataset}\label{subsec:dataset}

The magnetic evolution of the ISM has been viewed through isolated observational windows in the past.
Pulsar measurements primarily trace the diffuse ionized medium ($n_{\rm H} \lesssim 1$ cm$^{-3}$), while Zeeman splitting observations focus on denser neutral clouds ($n_{\rm H} \gtrsim 10^2$ cm$^{-3}$). 
This segregation, imposed by distinct observational techniques rather than physics, naturally fostered a segmented view of the ecosystem, where the ``broken power-law'' emerged as the most parsimonious description of the sparse data available at the time \citep{2010ApJ...725..466C}. 
Here, following \citet{2025MNRAS.tmp..512S}, we bridge this observational divide by assembling a unified dataset spanning ten orders of magnitude in density ($10^{-26}-10^{-16}$ g cm$^{-3}$) (Fig.\,\ref{figvisual}a). 
We integrated pulsar dispersion measures, applying established density calibrations \citep{2005AJ....129.1993M, 2017ApJ...850....4H, 2017ARA&A..55..111H}, with Zeeman data from H\textsc{i}, OH, and CN surveys \citep{2004ApJS..151..271H, 2008A&A...487..247F, 2010ApJ...725..466C, 2020ApJ...890..153J}. 
Recent analyses combining these observations with high-resolution multiphase MHD simulations confirm that turbulent cascades couple these thermodynamic phases into a globally consistent energy state, validating this joint analysis as a robust probe of the ISM's magnetic equation of state \citep{2022MNRAS.514..957S,2025MNRAS.tmp..512S}.
Technical details of the catalogue cuts and the density calibrations are given in Appendix\,\ref{ap.data}.

\subsection{Phenomenology: a gradual Steepening Slope}\label{subsec:phenom}

\begin{figure}[h]
    \centering
    \includegraphics[width=0.95\linewidth]{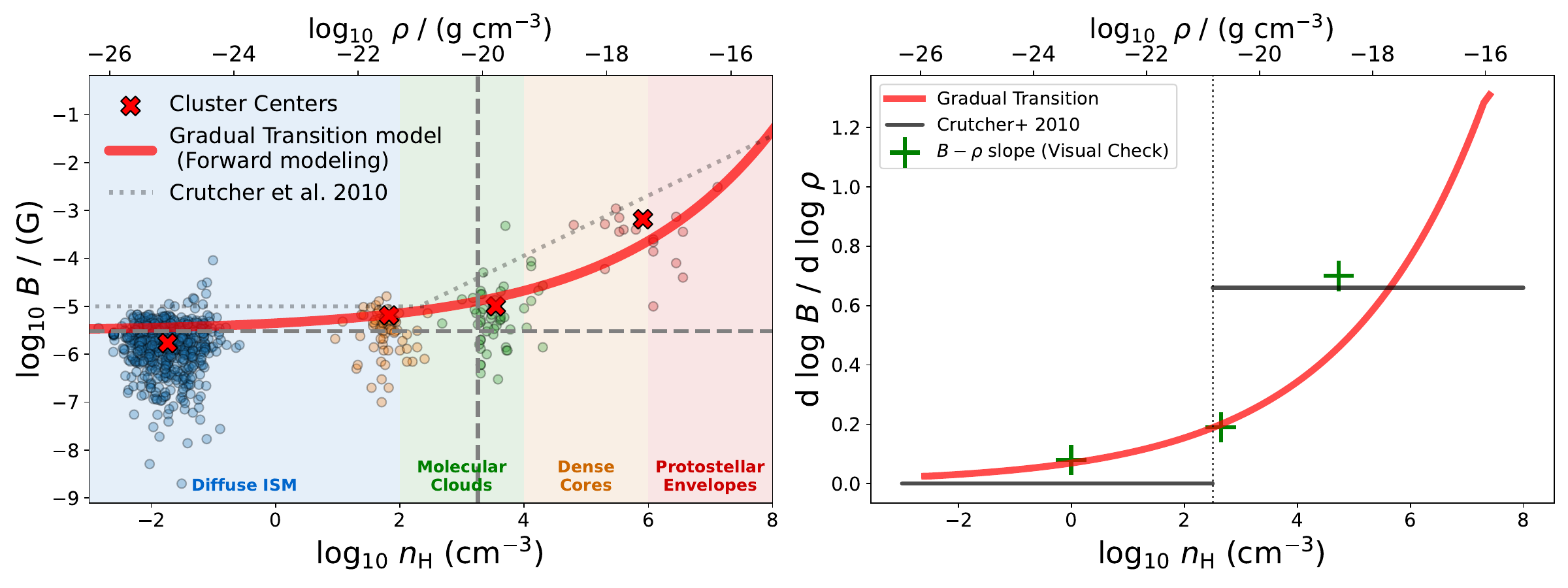}
    \caption{{\bf Continuous magnetic field evolution across interstellar scales.}
    Left panel, The magnetic field strength ($B_{\rm los}$) plotted against gas density ($\rho$) spanning over ten orders of magnitude. 
    Both tracers report the line-of-sight field; following the statistical isotropic expectation for randomly oriented fields \citep{2004ApJ...600..279C,2010ApJ...725..466C}, we display the observed B-field strength $B_{\rm los}$ to place the pulsar and Zeeman datasets on a common footing (Section\,\ref{subsec:forward}). 
    Observational data (scatter points) are compiled from pulsar measurements (diffuse ISM) and Zeeman splitting observations (molecular clouds to protostellar envelopes). 
    The background shading marks four density intervals, bounded at $n_{\rm H}\approx10^{2}$, $10^{4}$, and $10^{6}$ cm$^{-3}$: the diffuse ISM ($n_{\rm H}\lesssim10^{2}$ cm$^{-3}$), molecular gas ($10^{2}$--$10^{4}$ cm$^{-3}$), dense cores ($10^{4}$--$10^{6}$ cm$^{-3}$), and protostellar envelopes ($n_{\rm H}\gtrsim10^{6}$ cm$^{-3}$). The shading denotes these intervals only and carries no other quantity; the same boundaries are used in Fig.\,\ref{figphy}.
    The Gradual Transition model (solid red curve) provides a continuous description of the $B-\rho$ relation, passing through the density-binned cluster centers (red crosses). 
    For comparison, the gray dotted and dashed lines represent the piecewise step-function model from \citet{2010ApJ...725..466C}, which assumes a sharp transition at $n_{\rm H} \approx 300\,{\rm cm}^{-3}$.
    Right panel, The $B$-$\rho$ slope of the magnetic field--density relation ($k_{B-\rho} = d \log B / d \log \rho$) as a function of density. 
    The red curve shows the analytical derivative of the Gradual Transition model, revealing a smooth increase in the $B$-$\rho$ slope. 
    The green crosses represent coarse slope estimates derived from the finite differences between the cluster centers in panel a, serving as a visual consistency check; they are consistent with a smooth slope evolution rather than the discontinuous jump predicted by the broken power law (black steps).
    }
    \label{figvisual}
\end{figure}

The large dynamic range of this unified dataset exposes what the broken power law leaves unexplained (Fig.\,\ref{figvisual}): a piecewise fit can follow the data, but its break is a feature of the assumed functional form, not of a physical mechanism.
Contrary to the bimodal broken power law, the addition of pulsar data does not merely add a low-density floor. It reveals a trajectory that is visually consistent with a smooth evolution from the diffuse ISM to dense star-forming gas. This motivates testing a description in which the steepening is derived rather than inserted as a sharp break. 
The Gradual Transition model \citep{2024ApJ...976..209Z} provides one: a monotonic steepening of the scaling relation, from near-flat to steep with increasing density, that follows from the fundamental $\mathcal{M}_{\rm A}$-$k_{B\text{-}\rho}$ relation of compressive MHD turbulence. 
The observed trajectory is consistent with this form (the quantitative comparison, including the cases this dataset cannot decide, is deferred to Section\,\ref{sec:tests}). 
This apparent coherence motivates the hypothesis that the diffuse and dense regimes can be described as successive parts of a single continuous track. Whether the current data require this interpretation is tested quantitatively in Section~\ref{sec:tests}.

A corresponding visual indication is provided by the estimated evolution of the $B$-$\rho$ slope, $k_{B-\rho} = d \log B / d \log \rho$ (Fig.\,\ref{figvisual}b). 
Instead of the sharp jump from $k_{B-\rho} \approx 0$ to $k_{B-\rho} = 2/3$ imposed by step-function models, the $B$-$\rho$ slope estimates (green crosses, Fig.\,\ref{figvisual}b) indicate a monotonic steepening. 
This smooth evolution is in the same direction as the softened broken power laws of \citet{2025MNRAS.540.2762W} and \citet{2026arXiv260306838W}, and is consistent with magnetic and kinetic energies remaining coupled across the sampled density range \citep{2024ApJ...976..209Z}. 
Whether this smoothness is genuinely required by the data, rather than an impression created by combining different tracers, is the question we address quantitatively in Section\,\ref{sec:tests}.

\begin{figure}[h]
    \centering
    \includegraphics[width=0.95\linewidth]{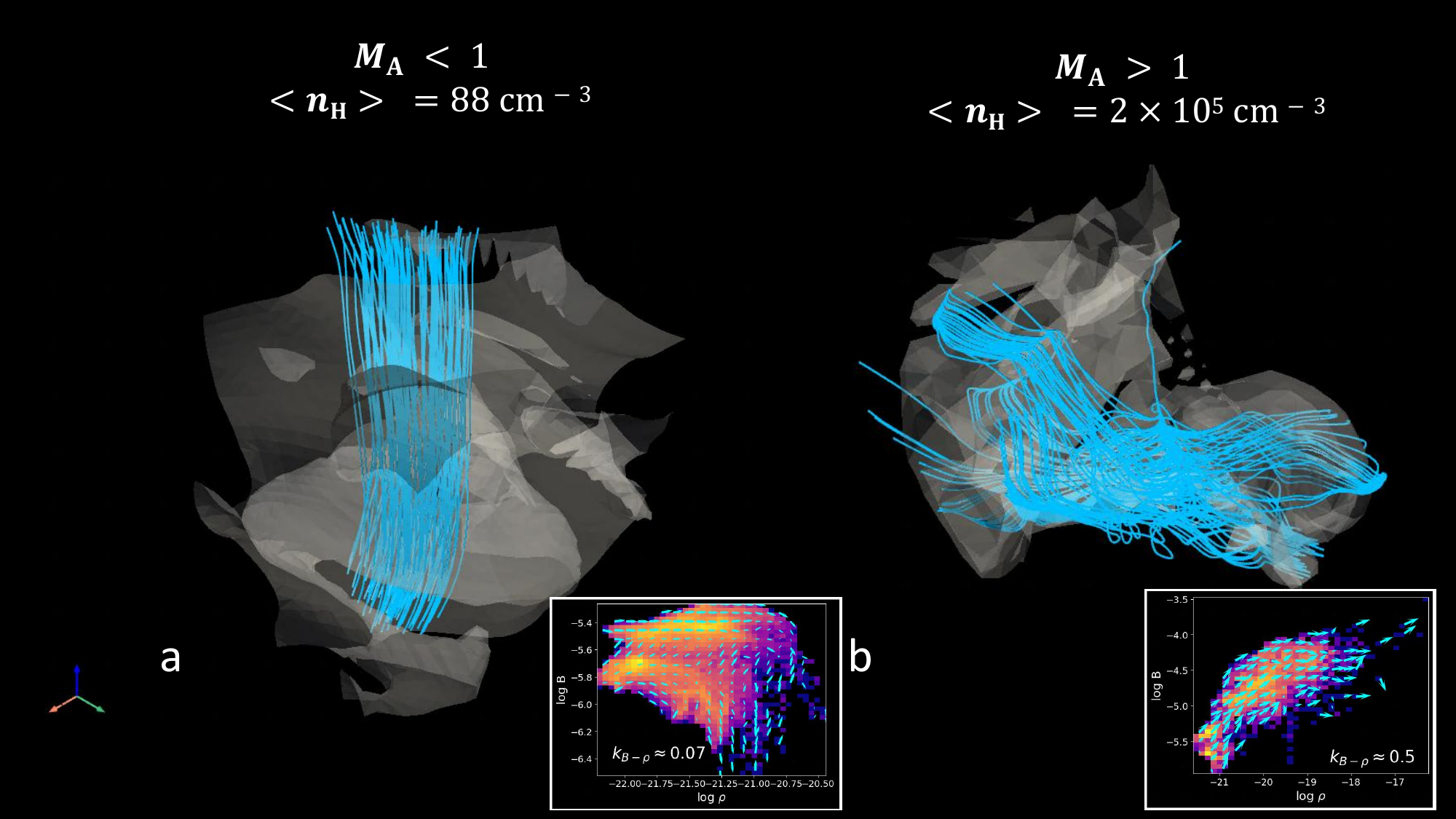}
    \caption{{\bf Morphological and statistical evidence of the magnetic transition.}
    Three-dimensional MHD simulations visualize the magnetic field structure (blue lines) and gas density (grey iso-surfaces) across the trans-Alfvénic threshold.
    a. In the sub-Alfvénic regime ($\mathcal{M}_A < 1$), magnetic field lines are ordered and rigid. 
    The corresponding phase-space analysis (inset) reveals a near-horizontal vector field with a slope $k_{B-\rho} \approx 0.07$.
    b. In the super-Alfvénic regime ($\mathcal{M}_A > 1$), the field lines become highly distorted and curved. 
    The inset shows the vector field steepening to $k_{B-\rho} \approx 0.5$. 
    The cyan arrows show the trend in phase space derived by Adjacent Correlation Analysis \citep{2025arXiv250605759L}.
    The visualizations represent local sub-volumes (32$^3$ and 16$^3$ voxels, respectively) extracted from the simulation to isolate the distinct Alfvénic regimes.
    }
    \label{figsim}
\end{figure}


\section{A Magnetic Equation of State based on $\mathcal{M}_{\rm A}$}\label{sec:eos}

\subsection{The Gradual Transition model}\label{subsec:gt}

The monotonic steepening of the $B-\rho$ relation reported in Section\,\ref{subsec:phenom} reflects a change in the energy balance of the gas.
The Gradual Transition model \citep{2024ApJ...976..209Z} identifies the Alfv\'en Mach number, $\mathcal{M}_{\rm A} = \sqrt{E_K/E_B}$, as the variable that organises the continuum between magnetically and kinetically dominated gas. 
The field strength follows
\begin{equation}\label{eq.GT}
    B(\rho) = B_c \exp\!\left[ \frac{1}{\beta} \left( \frac{\rho}{\rho_c} \right)^{\!\beta} \right],
\end{equation}
where $B_c$ is the background magnetic floor, $\rho_c$ the characteristic density at which the $B$-$\rho$ slope reaches unity ($k_{B-\rho}=1$), and $\beta = \gamma/\mathcal{K}$ the exponent set by the fundamental relation $\mathcal{M}_{\rm A} \propto k_{B-\rho}^{\mathcal{K}}$ together with the empirical scaling $\mathcal{M}_{\rm A} \propto \rho^{\gamma}$ (Appendix\,\ref{ap.forward}). 
The $B$-$\rho$ slope $k_{B-\rho} = d\log B/d\log\rho = (\rho/\rho_c)^\beta$ rises monotonically with density, so a single function spans the near-flat diffuse regime and the steep collapse regime. 
Because the slope passes continuously through every intermediate value, within the GT framework, the single power-law exponents reported in previous work can be interpreted as local slopes sampled over different density intervals. The Gradual Transition contains these power-law descriptions as its local tangents.

\begin{figure}[h]
    \centering
    \includegraphics[width=0.7\linewidth]{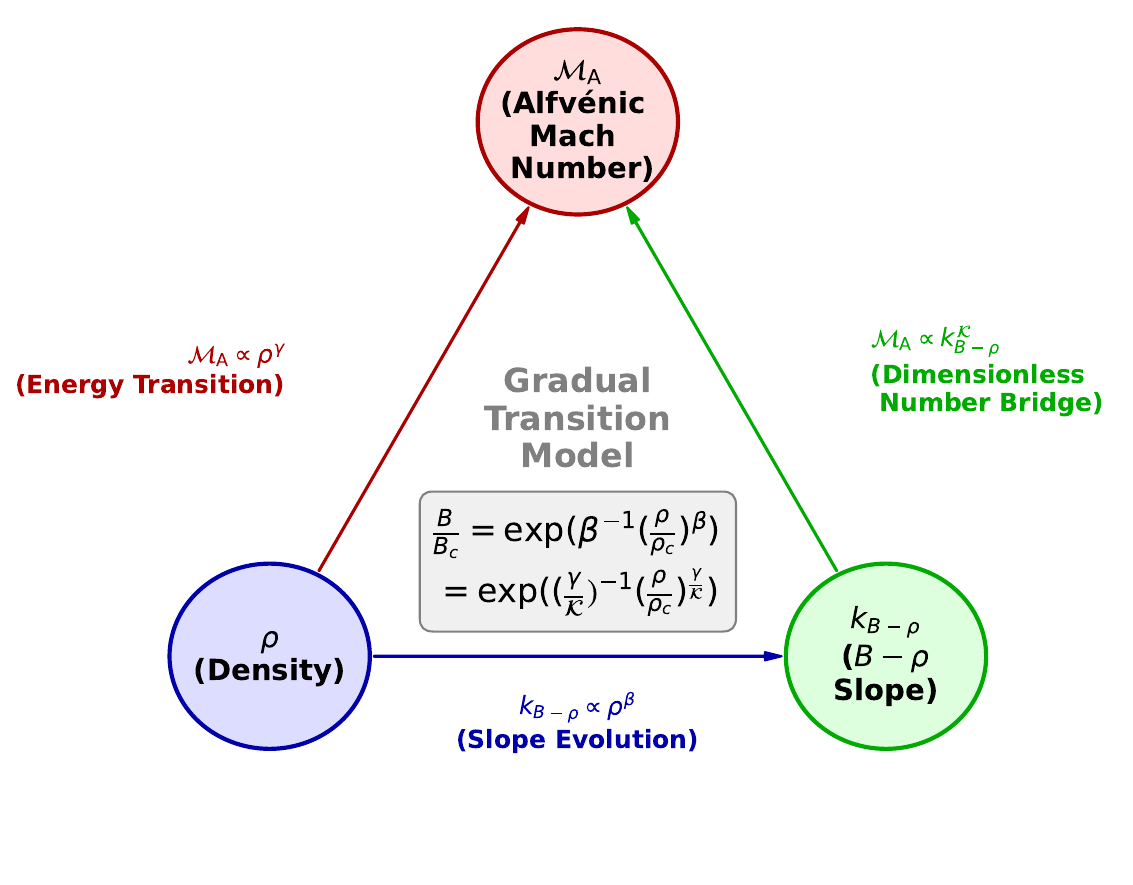}
    \caption{{\bf Logic of the Gradual Transition model.}
    The diagram links three quantities governing the magnetic evolution of the ISM: the gas density ($\rho$), the $B$-$\rho$ slope of the $B-\rho$ relation ($k_{B-\rho}$), and the Alfvén Mach number ($\mathcal{M}_{\rm A}$). 
    The bottom edge is the observational phenomenology: the relation steepens with density, characterised by the exponent $\beta$. 
    The right edge is the fundamental relation of compressive MHD turbulence \citep{2024ApJ...976..209Z}, which maps the geometric slope $k_{B-\rho}$ to the dynamical state $\mathcal{M}_{\rm A}$ via the scaling index $\mathcal{K} \approx 1.7$; this relation is used in Section\,\ref{subsec:gt}. 
    The left edge is the energy evolution: $\mathcal{M}_{\rm A}$ rises continuously with density ($\mathcal{M}_{\rm A} \propto \rho^\gamma$), marking the shift from magnetic regulation to gravity-driven turbulence. 
    The central box gives the self-consistency condition $\gamma = \beta\,\mathcal{K}$.}
    \label{figD3}
\end{figure}

The Alfv\'en Mach number is inferred from the $B$-$\rho$ slope through the fundamental relation $\mathcal{M}_{\rm A} \propto k_{B-\rho}^{\mathcal{K}}$ ($\mathcal{K} \approx 1.7$), calibrated in MHD simulations \citep{2024ApJ...976..209Z} in which $\mathcal{M}_{\rm A} = \sqrt{E_K/E_B}$ is measured directly from the fields, with $E_K = \tfrac{1}{2}\rho\,\sigma_v^2$ evaluated from the turbulent velocity dispersion on the scale at which density structure concentrates (Appendix\,\ref{ap.forward}). 
The turbulent velocity that defines $\mathcal{M}_{\rm A}$ therefore enters through the calibration: rather than measure a dispersion along each sightline, we adopt the slope-to-$\mathcal{M}_{\rm A}$ mapping that the simulated turbulence sets. 
Inference runs from the observed geometry ($k_{B-\rho}$) to the dynamical state ($\mathcal{M}_{\rm A}$), and from there to the energy ratio $E_K/E_B = \mathcal{M}_{\rm A}^2$. The inferred $E_K$ is not an independent observational quantity.
Accordingly, the observational values of $\mathcal{M}_{\rm A}$ quoted below are indirect, calibration-based inferences from the $B-\rho$ slope, rather than direct measurements from velocity dispersions. 
The shape of $B(\rho)$ is thus not a free function but is set by how $\mathcal{M}_{\rm A}$ evolves with density, which is what fixes the model's parameters in advance of the fit.

\subsection{Differentiated field behaviour across the transition}\label{subsec:morph}

$\mathcal{M}_{\rm A}$ organises the $B-\rho$ relation and its signature should be visible in the field itself.
In the observations $\mathcal{M}_{\rm A}$ is inferred; in the MHD simulation it is computed directly, so the simulation provides a test of whether $\mathcal{M}_{\rm A}$ governs the field geometry and the $B$-$\rho$ slope together. 
We compare two sub-volumes, one sub-Alfv\'enic ($\mathcal{M}_{\rm A}<1$) and one super-Alfv\'enic ($\mathcal{M}_{\rm A}>1$).

In the sub-Alfv\'enic gas ($\langle n_{\rm H}\rangle\approx88$ cm$^{-3}$) the magnetic energy exceeds the kinetic, and the field is ordered and rigid (Fig.\,\ref{figsim}a): the surrounding flows are less effective at bending it, and the gas is channelled along the lines without compressing them. Yet the adjacent-correlation vectors of the $B-\rho$ plane (cyan, Fig.\,\ref{figsim}a inset) are individually scattered, averaging to a near-flat slope $k_{B-\rho}\approx0.07$.
Magnetic field is too rigid for density fluctuations to move it systematically, so $B$ responds to $\rho$ incoherently and the net coupling is weak. 
In the super-Alfv\'enic gas ($\langle n_{\rm H}\rangle\approx2\times10^{5}$ cm$^{-3}$) the balance is reversed: the kinetic energy dominates and the field is bent and curved (Fig.\,\ref{figsim}b), but the correlation vectors now align point by point about a common slope $k_{B-\rho}\approx0.5$, because compression drives $B$ and $\rho$ together and the response is coherent. 
The two states therefore exhibit contrasting behaviours. In the sub-Alfv\'enic sub-volume, the magnetic field is morphologically ordered, while the local $B-\rho$ vectors are weakly aligned and yield only a small mean slope. In the super-Alfv\'enic sub-volume, the field morphology is more strongly distorted, while the $B-\rho$ vectors become more coherently aligned around a steeper slope. In these simulation sub-volumes, $\mathcal{M}_{\rm A}$ therefore organises both the magnetic-field morphology and the coherence of the field--density response, providing a behavioural manifestation of the relation $\mathcal{M}_{\rm A}\propto k_{B-\rho}^{\mathcal{K}}$.

The same transition appears in the observed field. The ordered diffuse geometry matches the parallel alignment between field and density structure seen by \emph{Planck} \citep{2013ApJ...774..128S,2016A&A...586A.138P}.
The curved geometry of the dense gas reproduces the hourglass morphology of star-forming regions \citep{2014ApJ...794L..18Q,2015Natur.520..518L,2017ApJ...846..122P} and the parallel-to-perpendicular flip of the \emph{Planck} relative-orientation statistics \citep{2016A&A...586A.138P,2017A&A...607A...2S}. 
The same organisation holds on intermediate scales: velocity-gradient mapping of Orion A recovers the molecular-cloud field structure and strength across scales and compares it with \emph{Planck} dust polarisation \citep{2022ApJ...934...45Z}, while in Galactic giant filaments the otherwise regular field is interrupted by local magnetization gaps where gravity, collapse, or feedback disturb it \citep{2024ApJ...961..124Z}.
This is evidence for the role of $\mathcal{M}_{\rm A}$ that does not pass through the $B-\rho$ fit. 
Because $\mathcal{M}_{\rm A}$ controls the field in this way and rises with density, the evolution from the diffuse ISM to collapsing cores naturally connects to the magnetic, kinetic, and transition regimes that are spatially segregated in compressive MHD turbulence \citep{2025MNRAS.542.3246L}.

\subsection{Three regimes and the role of magnetic fields}\label{subsec:regimes}

The Gradual Transition (Equation\,\ref{eq.GT}) contains three regimes and fixes its parameters from independent physics rather than from a fit to the present data.
The consistency of the resulting externally calibrated forward curve with the data is tested in Section\,\ref{sec:tests}.

\paragraph{Galactic magnetic floor.}
As $\rho\to0$ the relation approaches the constant floor $B_c$, which in the Milky Way is the background field of the diffuse ISM, $B_c\approx2.0\,\mu$G \citep{2006ApJ...642..868H,2017ARA&A..55..111H}. 
The corresponding magnetic pressure, $B_c^2/8\pi\approx1.2\times10^3$ K cm$^{-3}$, is of the same order as the thermal pressure of the Warm Neutral Medium \citep[$\approx3\times10^3$ K cm$^{-3}$;][]{2003ApJ...587..278W}, confirming that a floor of this strength is energetically reasonable for the diffuse gas. 
That the gas is sub-Alfv\'enic ($\mathcal{M}_{\rm A}<1$, $E_B>E_K$) follows from its near-flat slope through the fundamental relation (Section\,\ref{subsec:gt}), and the near-flat scaling $k_{B-\rho}<0.2$ is the signature of this regime.

\paragraph{Trans-Alfv\'enic transition.}
We identify $\mathcal{M}_{\rm A}=1$ as the trans-Alfv\'enic equipartition point and associate this transition with the increasing contribution of gravitational energy injection. We anchor it at $n_{\rm H}\simeq1800$ cm$^{-3}$, where independent simulations place the onset of significant gravitational injection into the turbulent cascade \citep{2025ApJ...988..132Y}. This association is an input to the forward model rather than an independent measurement from the present $B-\rho$ dataset. 
The break density of broken power laws is not sharply defined: as the density uncertainties are treated as a free parameter and the slope bound is widened, the estimated $n_0$ drifts upward, from $\approx300$ cm$^{-3}$ \citep{2010ApJ...725..466C} to $\sim1100$ cm$^{-3}$ \citep{2020ApJ...890..153J}. 
This upward drift is qualitatively consistent with the Gradual Transition interpretation. In the GT model there is no intrinsically sharp break; instead, a broken-power-law fit approximates a continuous transition, and its fitted break density can shift as the adopted uncertainty model and slope constraints are varied.

\paragraph{Gravitational collapse and magnetic regulation.}
Beyond equipartition ($\mathcal{M}_{\rm A}>1$) the magnetic field no
longer dominates the energy budget. One expectation is $k_{B-\rho}=2/3$ for weak-field isotropic spherical contraction \citep{1966MNRAS.133..265M}, whereas magnetically regulated anisotropic contraction can produce shallower scalings around $k_{B-\rho}\simeq1/2$ \citep{1999ASIC..540..305M,2015MNRAS.451.4384T}.
The model gives no such jump: the slope $(\rho/\rho_c)^\beta$ rises
continuously through intermediate values. Magnetic tension continues
to set the geometry of collapse after it ceases to dominate the energy
\citep{2024arXiv240809690Z}, holding the slope below the isotropic
value \citep{1999ASIC..540..305M}, so the exit from magnetic
regulation is as gradual as the entry. The historical exponents then
mark densities rather than regimes. For the adopted parameters the
slope reaches $1/2$ near $n_{\rm H}\sim7\times10^{4}\,{\rm cm^{-3}}$,
in the starless-core window where \citet{2015MNRAS.451.4384T} recover
$k_{B-\rho}\approx1/2$ from the Crutcher data with relaxed density
uncertainties, and reaches $2/3$ only near
$n_{\rm H}\sim4\times10^{5}\,{\rm cm^{-3}}$, among the
strongest-field detections in the sample. Within the GT interpretation, each historical exponent corresponds to the local slope reached by the same continuous relation at a different density, and
each lies below the isotropic value until well past the densities the
debate has been fought over.
Observations of dense star-forming gas bear out this dual role: in the massive star-forming complex M17\,SW, high-resolution dust polarisation resolves the magnetic-field structure and the relative magnetic, kinetic, and gravitational energies, identifying regions that are gravity-dominated alongside regions still regulated by the magnetic field \citep{2026ApJ..1005...91Z}.

These three regimes are segments of one gradual-transition relation (Eq.\,\ref{eq.GT}), not separate laws. The gas passes through them continuously as it evolves from the diffuse ISM to star-forming cores.

\section{Confrontation with the data}\label{sec:tests}

The GT model has three parameters, each fixed from independent physics before any contact with the present dataset (Section\,\ref{subsec:gt}; Appendix\,\ref{ap.forward}). This section asks whether the data support those values, and how far the Gradual Transition can be pushed before it fails. The Bayesian comparison turns out to be inconclusive on the present sample; the more informative results come from the parameter recovery and from the blind extrapolation.

\subsection{Forward model}\label{subsec:forward}

The background field $B_c \approx 2.0~\mu$G is the coherent large-scale field from pulsar rotation measures \citep{2006ApJ...642..868H,2017ARA&A..55..111H}. The scaling exponent $\beta = \gamma/\mathcal{K} \approx 0.147$ combines the simulation-calibrated fundamental relation ($\mathcal{K} \approx 1.7$; \citealt{2024ApJ...976..209Z}) with the empirical density scaling ($\gamma \approx 0.25$; Appendix\,\ref{ap.forward}). 
The characteristic density $\rho_c$ is derived by anchoring $\mathcal{M}_{\rm A}=1$ at $n_{\rm trans}\approx1800$ cm$^{-3}$ \citep{2025ApJ...988..132Y} and propagating along the same scaling ($\mathcal{M}_{\rm A} \propto\rho^{0.25}$; Appendix\,\ref{ap.forward}); of the three parameters it is the least tightly constrained, so we treat it as anchored rather than recovered (Section\,\ref{subsec:bic}).

\begin{figure}[h]
    \centering
    \includegraphics[width=0.8\linewidth]{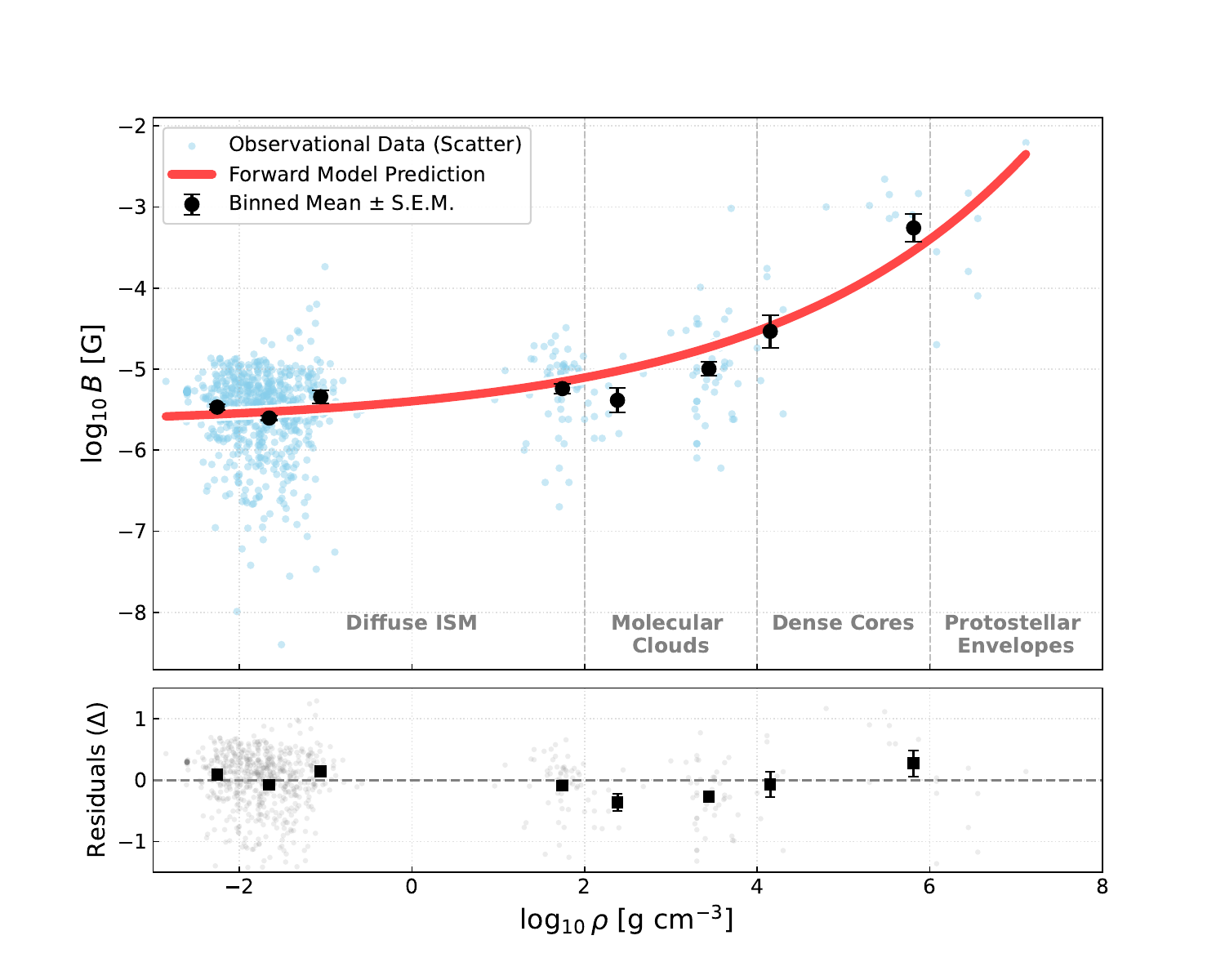}
    \caption{{\bf Forward-model test of the magnetic field–density relation.} 
    a, Comparison between the theoretical prediction and observational data. The solid red curve represents the Gradual Transition model ($B_c = 2.0\,\mu\text{G}$, $n_{\text{trans}} \approx 1800\,\text{cm}^{-3}$), which closely follows the binned mean magnetic field strengths (black circles; error bars denote s.e.m.) derived from the individual measurements (light blue scatter). 
    The background shading marks the same density regimes as in Fig.\,\ref{figvisual} (boundaries at $n_{\rm H}\approx10^{2}$, $10^{4}$, $10^{6}$ cm$^{-3}$).
    b, Residuals of the logarithmic magnetic field strength ($\Delta = \log_{10} B_{\rm obs} - \log_{10} B_{\rm model}$) as a function of gas density. 
    The binned residuals (black squares) align with the zero-deviation line (dashed), showing that the binned residuals exhibit no clear density-dependent systematic offset.
    The residual distribution in the whole region is shown in Fig.\,\ref{figD2}a.}
    \label{figphy}
\end{figure}

With no parameter adjusted to the data, the resulting curve passes through the density-binned means across all ten decades (Fig.\,\ref{figphy}a). The mean residual is close to zero ($\langle\Delta\log B\rangle \approx -0.11$ dex, $\sigma\approx0.68$ dex), and the binned residuals show no obvious systematic trend with density (Fig.\,\ref{figphy}b; Appendix\,\ref{ApB2}). The scatter is large because the field at fixed density is spread by the distribution of $\mathcal{M}_{\rm A}$ among sightlines (Appendix\,\ref{ap.scatter}); the forward model predicts the median, not each line of sight. The agreement is a prediction, not a fit.

\subsection{Bayesian model comparison}\label{subsec:bic}

We compare the GT model with the generalised four-parameter BPL ($\alpha_1$ free; \citealt{2020ApJ...890..153J, 2026arXiv260306838W}) using $\Delta\text{BIC} = \text{BIC}_{\rm BPL} - \text{BIC}_{\rm GT}$, under two density-error assumptions ($R=2$, $R=9.3$) and two likelihoods (flat-envelope and symmetric lognormal-median; Appendix\,\ref{Ap.BIC}). 
The verdict is sensitive to these choices.
An injection-recovery test shows that the flat-envelope likelihood does not reproduce the observed data well, so preferences computed under it should be read with care (Appendix\,\ref{ap.inject}).

The dense gas alone cannot separate the two forms. The three Zeeman tracers sample disconnected density windows, and within any single window a smooth law is locally indistinguishable from a power law. 
The two descriptions diverge most where the sampling is sparsest, in the H\textsc{i}, OH gap at $n_{\rm H}\sim300$--$500$ cm$^{-3}$. 
This limitation is symmetric, it leaves the break of the broken power law weakly determined just as it prevents the Gradual Transition from being established, and is removed only by new measurements in the gap, not by reanalysis. 
On the unified dataset the sign of $\Delta\text{BIC}$ instead depends on how the cross-tracer projection factor $f_p$ is modelled, spanning $+4.3$ to $-8.5$ with no specification reaching the decisive threshold $|\Delta\text{BIC}|>10$ \citep{1995JASA...90..773K}.
The functional form is statistically underdetermined (Table\,\ref{tab:bic_comparison}).

What the fits do show is that the GT fits return parameter values consistent with the externally specified predictions (Table\,\ref{tab:predictions}). 
The exponent settles at $\beta=0.15$--$0.21$ around the predicted $0.147$, the floor at $B_c=1.9$--$2.3\,\mu$G around the adopted $2.0\,\mu$G, and $f_p$ within its expected range.
The broken-power-law fits, by contrast, leave the diffuse slope at $\alpha_1=0$ and the break density floating with the assumed density errors, so their parameters are not set in advance. 
An unconstrained MCMC fit\label{subsec:posterior} returns posteriors consistent with the constrained values (Fig.\,\ref{figD4}, Table\,\ref{tab.params}), with a moderate $\rho_c$--$\beta$ correlation that broadens the marginal constraint on $\rho_c$, while leaving the posterior predictive $B(\rho)$ relation better constrained over the observed density range; $\rho_c$ is the most likelihood-sensitive parameter, shifting by about a decade, which is why we treat it as anchored rather than recovered. 
The case for the Gradual Transition model therefore rests on this parameter-level agreement, on the blind extrapolation below, and on the spectral prediction of Section\,\ref{sec:prediction}, rather than on a model-selection statistic.

\subsection{Blind extrapolation}\label{subsec:blind}

\begin{figure}[h]
    \centering
    \includegraphics[width=0.8\linewidth]{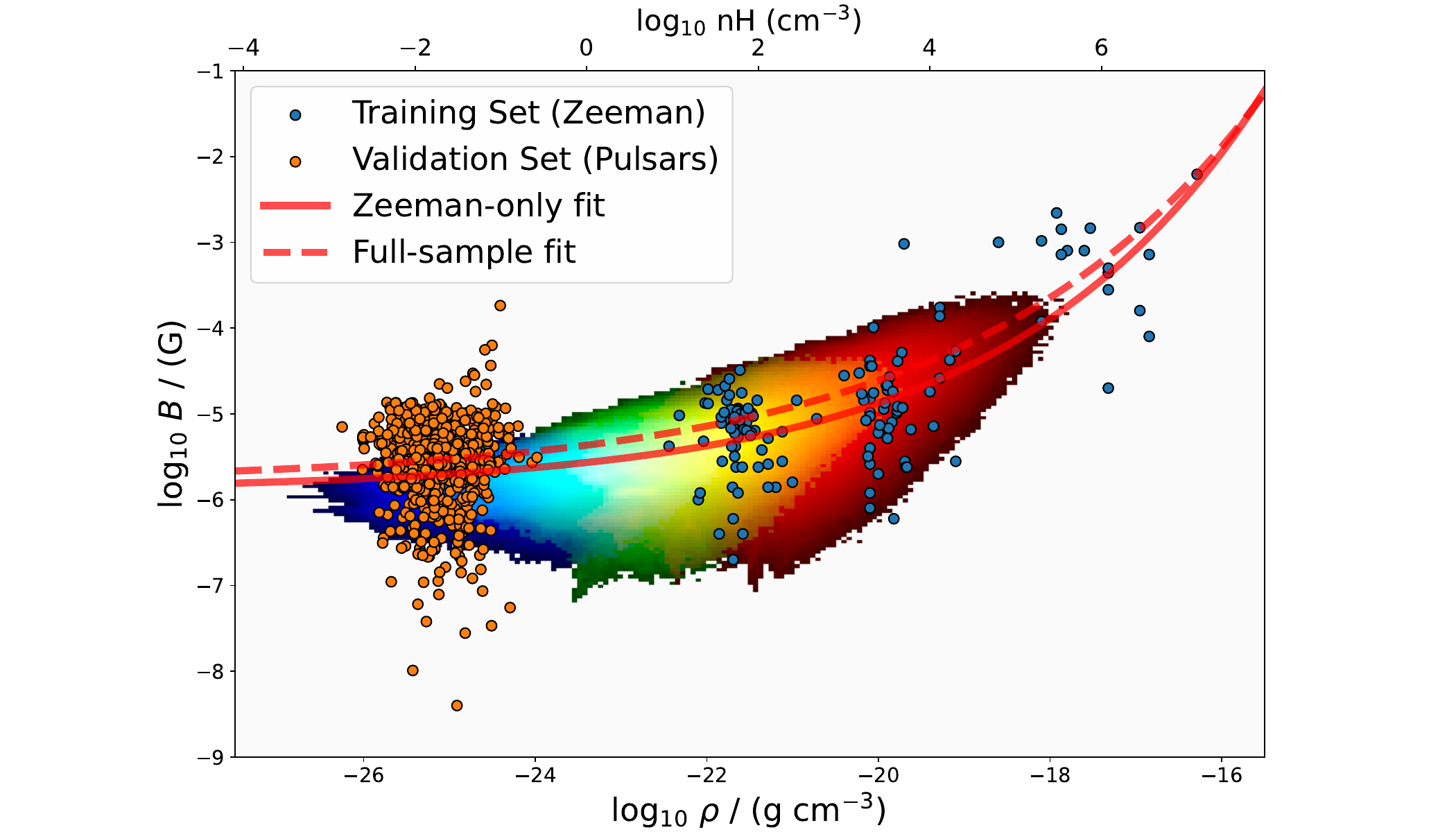}
    \caption{{\bf Blind extrapolation test.}
    The Gradual Transition model fitted to the full unified dataset (dashed red line) is compared with the same model fitted only to the dense-gas Zeeman sample (solid red line, $n_{\rm H} \gtrsim 100$ cm$^{-3}$), with the diffuse pulsar data held out during fitting. 
    The dense-only fit extrapolated over four decades in density passes through the diffuse-ISM pulsar measurements, showing that the agreement with the diffuse pulsar regime is not solely a consequence of including the pulsar measurements in the fit. 
    All measurements are shown on the ensemble-mean total-field scale, $B_{\rm tot}\simeq2|B_{\rm los}|$, consistent with the convention adopted for these fits (Appendices~\ref{ApB2} and~\ref{apc8}). 
    The background grey scatter shows the $B-\rho$ distribution from MHD simulations at three values of $\mathcal{M}_{\rm A}$ \citep{2024ApJ...976..209Z} and is included for reference.}
    \label{figE2}
\end{figure}

We fit the GT model to the Zeeman data alone and extrapolate into the diffuse regime without adjustment (Fig.\,\ref{figE2}), fitting the median of the distribution rather than the upper envelope $B_{\rm max}(\rho)$ (Appendix\,\ref{ap.inject}). 
The dense-only and full-sample fits are nearly indistinguishable: the diffuse-end prediction is already encoded in the dense gas. 
The pulsar measurements, which played no part in the calibration, fall on the extrapolated curve within the population scatter across four decades in density.

The Bayesian comparison does not decide between the functional forms, but the three lines of evidence provide complementary support: a continuous $B(\rho)$ whose parameters are predicted in advance, whose shape survives blind extrapolation, and whose spectral consequence is testable. A model-independent estimate of the $B$-$\rho$ slope, free of any global fit and consistent with this picture, is given in Appendix\,\ref{ap.local}.

\begin{figure}[h]
    \centering
    \includegraphics[width=0.85\linewidth]{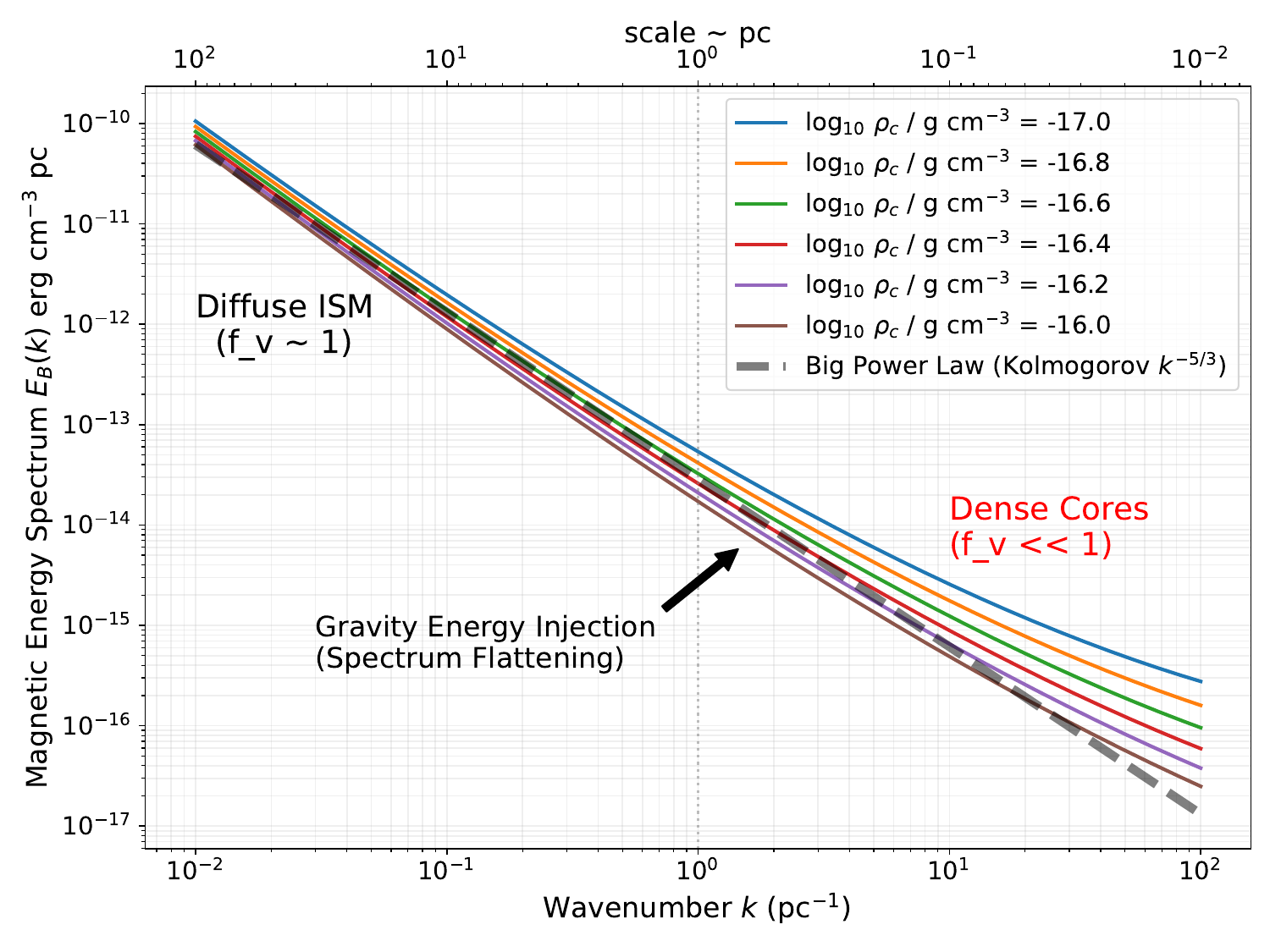}
    \caption{{\bf Magnetic-energy spectrum implied by the Gradual Transition model under the adopted scale mapping.}
    The continuous $B(\rho)$ relation predicts a magnetic energy spectrum $E_B(k)$ (solid coloured lines, computed from Eq.\,\eqref{eq.spectrum}) that follows the Kolmogorov ``Big Power Law'' of electron-density fluctuations (black dashed line, $E \propto k^{-5/3}$) \citep{1995ApJ...443..209A,2010ApJ...710..853C} at large scales and flattens at small scales as gravitational compression steepens $B(\rho)$. 
    The colour gradient corresponds to different characteristic densities $\rho_c$ spanning the MCMC posterior (Table\,\ref{tab.params}); the location of the flattening shifts with $\rho_c$ while the qualitative shape persists. 
    The adopted size--density relation $\rho \propto r^{-1.1}$ \citep{1981MNRAS.194..809L} implies $D_f \simeq 1.9$ and hence $f_v(k) \propto k^{D_f-3} \simeq k^{-1.1}$ (Appendix\,\ref{ap.spectrum}); because these inputs are specified by the adopted Larson relation rather than fitted to the magnetic-field data, the curves constitute a conditional prediction of the model under the adopted scale mapping.}
    \label{figD5}
\end{figure}

\section{The Magnetic Energy Spectrum}\label{sec:prediction}

The continuous $B(\rho)$ makes a prediction beyond the $B-\rho$ plane itself. 
Previous work derived magnetic-energy spectra from cloud density statistics and the B-$\rho$ relation, and found approximate multiscale magnetic--gravitational equipartition in NGC 6334 \citep{2018MNRAS.474.2167L}.
Here, we map the continuous GT relation into wavenumber space. Under the adopted size--density relation and volume-filling-factor prescription, the continuous $B(\rho)$ relation implies a phenomenological magnetic-energy spectrum $E_B(k)$, with no additional parameter fitted to the magnetic-field dataset.

The resulting spectrum has two regimes (Fig.\,\ref{figD5}). 
At large scales ($k\lesssim 1$ pc$^{-1}$) the field sits near its floor $B\approx B_c$, and $E_B(k)$ follows the Kolmogorov $k^{-5/3}$ form of the ``Big Power Law'', the spectrum of electron-density fluctuations spanning many decades in the diffuse Galactic ISM \citep{1995ApJ...443..209A,2010ApJ...710..853C}. 
Under this mapping, the Gradual Transition spectrum approaches a $k^{-5/3}$-like large-scale scaling consistent with the Galactic Big Power Law. 
At small scales ($k\gtrsim 1$ pc$^{-1}$, $n_{\rm H}\gtrsim10^3$ cm$^{-3}$) gravity compresses the gas and amplifies the field, lifting $E_B(k)$ above the Kolmogorov extrapolation: the spectrum flattens. 
This onset falls at the characteristic scale and density of self-gravitating molecular clumps \citep{1987ApJ...319..730S, 2007ARA&A..45..339B}, where gravity starts to bind interstellar structure, the same transition that the $B-\rho$ relation marks in density space, now seen in wavenumber space. 
The flattening wavenumber is fixed by the same $\rho_c$ that governs the $B-\rho$ relation. It shifts with $\rho_c$ (Fig.\,\ref{figD5}) but the feature itself persists, because it is driven by the $B(\rho)^2$ amplification at high density.

The physical content is a change in the role of the magnetic field across the spectrum. In the diffuse gas the field follows the turbulent cascade near-passively. 
In dense gas gravity drives the kinetic energy that amplifies the field above the cascade. A measurement of $E_B(k)$ in dense filaments and cores tests this directly: the Gradual Transition model predicts a small-scale excess over the Big Power Law extrapolation, growing with density.

\section{A continuous magnetic equation of state}\label{sec:discussion}

We have assembled a $B$--$\rho$ dataset spanning ten orders of magnitude in density, from pulsar sightlines through the diffuse ISM to Zeeman measurements in dense cores, and tested against it the Gradual Transition model, in which the GT model uses $\mathcal{M}_{\rm A}$ to organise the field--density relation at each density. Fits to the data recover parameter values consistent with those specified in advance from independent physical constraints. A dense-gas fit extrapolates blindly across four decades onto the pulsar measurements. Under the adopted scale mapping, the implied magnetic-energy spectrum approaches a $k^{-5/3}$-like large-scale scaling and departs from that extrapolation at small scales.

The Gradual Transition model does not require the broken-power-law phenomenology to be discarded: over restricted density intervals, the two asymptotic parts of the continuous relation can be approximated by the two branches of a broken power law. The empirical break density $n_0$ may be interpreted as an effective approximation to the broader transition around $\mathcal{M}_{\rm A}=1$, and the two magnetic regimes of the broken power law are the two ends of one equation of state. Independent work points in the same direction: the energy-density proportionality of \citet{2025MNRAS.tmp..512S}, the softened broken power laws of \citet{2025MNRAS.540.2762W} and \citet{2026arXiv260306838W}, and the field morphology changing continuously from ordered to curved across the transition (Fig.\,\ref{figsim}). What distinguishes the Gradual Transition model from the broken power law is not a statistical preference on the present data, which cannot decide the functional form, but that its parameters are derived and predicted rather than fitted, and that it reaches into observables (the scatter structure and the energy spectrum) that the broken power law does not address.

The slope-to-$\mathcal{M}_{\rm A}$ calibration rests on ideal, isothermal simulations \citep{2024ApJ...976..209Z}, whereas the sightlines integrate over up to $\sim10$ kpc of multiphase gas; $\mathcal{M}_{\rm A}$ is inferred from the slope rather than measured from velocities; and the pulsar $B_\parallel$ assumes uncorrelated electron density and field along the sightline \citep{2021MNRAS.502.2220S}. These limitations affect both the inferred parameter values and the strength of the physical interpretation connecting the observed slope to $\mathcal{M}_{\rm A}$. They do not, however, remove the empirical motivation for testing a continuous $B(\rho)$ description.

Within the GT interpretation, the diffuse and dense ISM represent two ends of a continuous magnetic evolution rather than fundamentally separate scaling regimes. In the diffuse gas the field traces the turbulent cascade; in dense cores gravity drives the kinetic energy that amplifies it, its imprint the spectral flattening at small scales. Star formation is then not an event that breaks the scaling law but a consequence of the changing role of gravity, connecting the physics of core collapse to the magnetised structure of the Galaxy across ten decades in density. 
On galaxy scales, related magnetic organisation is also seen: the field of NGC\,628 follows the spiral arms and bends around feedback-driven bubbles, tracing structure on kiloparsec scales \citep{2024ApJ...967...18Z}.

\begin{acknowledgments}

We thank the referee for helpful comments that improved the clarity of the manuscript.

M.Z. acknowledges support from the National Natural Science Foundation of China (NSFC) under grant No. 12503029.
G.-X.L. acknowledges support from the NSFC under grant No. 12273032.
K.Q. is supported by the NSFC under grant No. 12425304, the National Key R\&D Program of China (Nos. 2023YFA1608204 and 2022YFA1603103), and the grant from the China Manned Space Project.

\end{acknowledgments}

\bibliographystyle{aasjournalv7}
\bibliography{reference}

\appendix

\setcounter{figure}{0}
\renewcommand{\thefigure}{A\arabic{figure}}
\renewcommand{\theHfigure}{A\arabic{figure}}

\renewcommand{\theHequation}{\thesection.\arabic{equation}}


\section{Data Compilation $\&$ Processing}\label{ap.data}

The unified dataset ($N=761$) comprises two components.
\subsection{Zeeman splitting data}
We used 137 high-confidence Zeeman measurements ($n_{\rm H} \gtrsim 10^2$ cm$^{-3}$) from H\textsc{i}, OH, and CN spectral lines. 
The sample draws from the Millennium Arecibo 21-cm Absorption-Line Survey \citep{2004ApJS..151..271H}, the Arecibo OH Absorption Survey \citep{2008ApJ...680..457T}, and IRAM 30m CN emission measurements \citep{2008A&A...487..247F}, as compiled by \citealt{2010ApJ...725..466C}. 
We adopt the line-of-sight field strengths ($B_{\rm los}$), volume densities, and uncertainties from the original catalogues.
\subsection{Pulsar dispersion measures}
To constrain the diffuse ionized medium ($n_{\rm H} \lesssim 1$ cm$^{-3}$), we selected 624 pulsars from the ATNF Pulsar Catalogue (v2.60) \citep{2005AJ....129.1993M}. 
Following the quality cuts of \citealt{2025MNRAS.tmp..512S}, we excluded sources with null measurements or zero uncertainty in Rotation Measure (RM) or Dispersion Measure (DM), and restricted the sample to sources within a path length of $L_{\rm dist} < 10$ kpc to minimize geometric depolarization effects.
The line-of-sight magnetic field was derived as:$$\langle B_{\parallel} \rangle = 1.232 \frac{\text{RM}}{\text{rad m}^{-2}} \left( \frac{\text{DM}}{\text{pc cm}^{-3}} \right)^{-1} \, \mu\text{G}.$$
This DM-weighted estimator assumes that the thermal electron density and the magnetic field are uncorrelated along the sightline; a correlation between the two biases $\langle B_{\parallel}\rangle$, the sign of the bias following the sign of the correlation \citep{2021MNRAS.502.2220S}. 
For the diffuse, turbulent gas traced by pulsars this correlation is weak, and the resulting bias is small compared with the order-of-magnitude dynamic range spanned here; the effect is tested in Appendix\,\ref{Ap.BIC}. 
To derive hydrogen number densities ($n_{\rm H}$) from the pulsar-derived average electron densities ($\langle n_e \rangle = \text{DM}/L_{\rm dist}$), we applied the ionization correction formulated by \citealt{2017ApJ...850....4H} to account for the warm ionized medium phase:$n_e \approx n_{\rm H} \left[ \frac{\arctan(T/10^3\,\text{K} - 10)}{\pi} + 0.5 \right],$
assuming a characteristic temperature $T = 10^4$ K. 
All densities are converted to mass density ($\rho = \mu m_p n_{\rm H}$) assuming a mean molecular weight per hydrogen molecular of $\mu = 2.8$.

\begin{figure}
    \centering
    \includegraphics[width=0.49\linewidth]{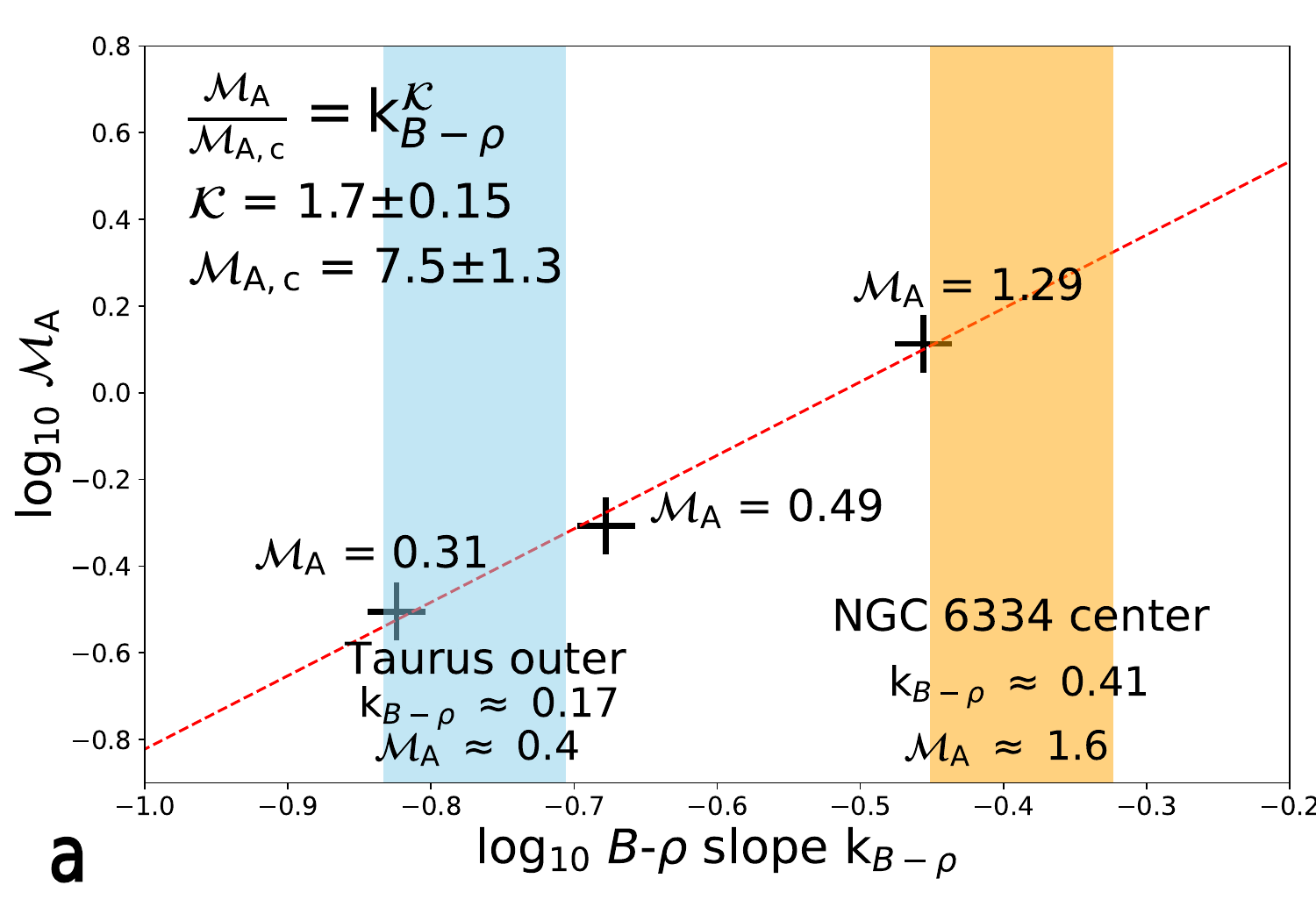}
    \includegraphics[width=0.46\linewidth]{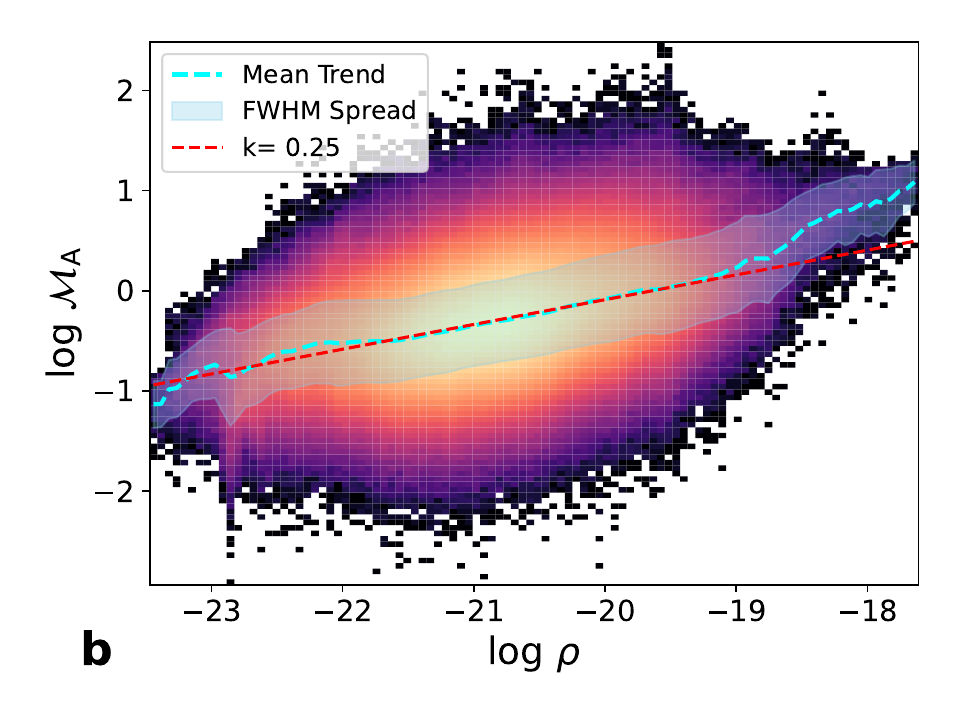}
    \caption{{\bf Theoretical scaling relations from MHD simulations.}
    left, The fundamental relation linking the local magnetic field–density slope ($k_{B-\rho} = d \log B / d \log \rho$) to the Alfvén Mach number ($\mathcal{M}_A$). The red dashed line shows the power-law fit $\mathcal{M}_A \propto k_{B-\rho}^{\mathcal{K}}$ with $\mathcal{K} \approx 1.7$, derived from the statistical analysis of MHD simulations (using the same simulation dataset as in b to ensure physical consistency). 
    right, The evolution of the Alfvén Mach number with gas density in the simulation. The data (density-colored scatter) reveal a power-law trend $\mathcal{M}_A \propto \rho^{\gamma}$ with $\gamma \approx 0.25$ (red dashed line). 
    The cyan dashed line traces the mean trend, and the shaded area indicates the FWHM spread. Combining these two scaling laws ($\beta = \gamma / \mathcal{K}$) yields the predicted slope evolution exponent $\beta \approx 0.147$ used in the forward model.}
    \label{figD1}
\end{figure}

\subsection{MHD Simulations}
The MHD simulation \citep{2024ApJ...976..209Z} uses the \textsc{Enzo} code \citep{2012ApJ...750...13C}, initialised with super-Alfvénic turbulence ($\mathcal{M}_s = 9$, $\beta_0 = 20$, $\alpha_{\rm vir} = 1$) in a $(4.6 \text{ pc})^3$ box with $256^3$ resolution, evolved with self-gravity to $0.6\,t_{\rm ff}$ ($0.76$ Myr) \citep{2024ApJ...976..209Z}. 
We compute $\mathcal{M}_{\rm A}$ on a scale of 8 pixels ($0.144$ pc), where the density power spectrum peaks \citep{2024ApJ...976..209Z}, and extract a $32^3$ sub-cube ($0.6$ pc) that is entirely sub-Alfv\'enic and a $16^3$ sub-cube ($0.3$ pc) that is entirely super-Alfv\'enic (Fig.~\ref{figsim}).
We derived the scaling laws ($\mathcal{M}_A \propto k_{B-\rho}^{1.7}$ \citep{2024ApJ...976..209Z} and $\mathcal{M}_A \propto \rho^{0.25}$) from the full simulation volume to determine the exponent $\beta$ (see Fig.\,\ref{figD1}).

\subsection{Applicability and Scatter}\label{ap.scatter}
The simulations are isothermal, ideal, and span a single molecular-cloud box, whereas the observations are line-of-sight averages, and the pulsar sightlines integrate over up to $\sim 10$ kpc of diffuse gas. 
These differ in scale and physics, so the simulations are not used to reproduce individual measurements. 
We use them only to calibrate the fundamental relation, $\mathcal{M}_{\rm A}(k_{B-\rho})$, and empirical relation,$\mathcal{M}_{\rm A}(\rho)$, which set the exponent $\beta$; the dimensional anchors $B_c$ and $\rho_c$ are fixed from observations and from the onset density of gravity-driven turbulence, not from the box. 
The relevant quantities are $B$-$\rho$ slopes and the energy ratio, both of which are dimensionless and less sensitive to the absolute scale than the field strength itself. 
Path-length averaging adds scatter to the inferred slope but, for orientations uncorrelated with density, does not bias the ensemble mean exponent (Appendix\,\ref{ap.local}).

\begin{figure}[h]
    \centering
    \includegraphics[width=0.8\linewidth]{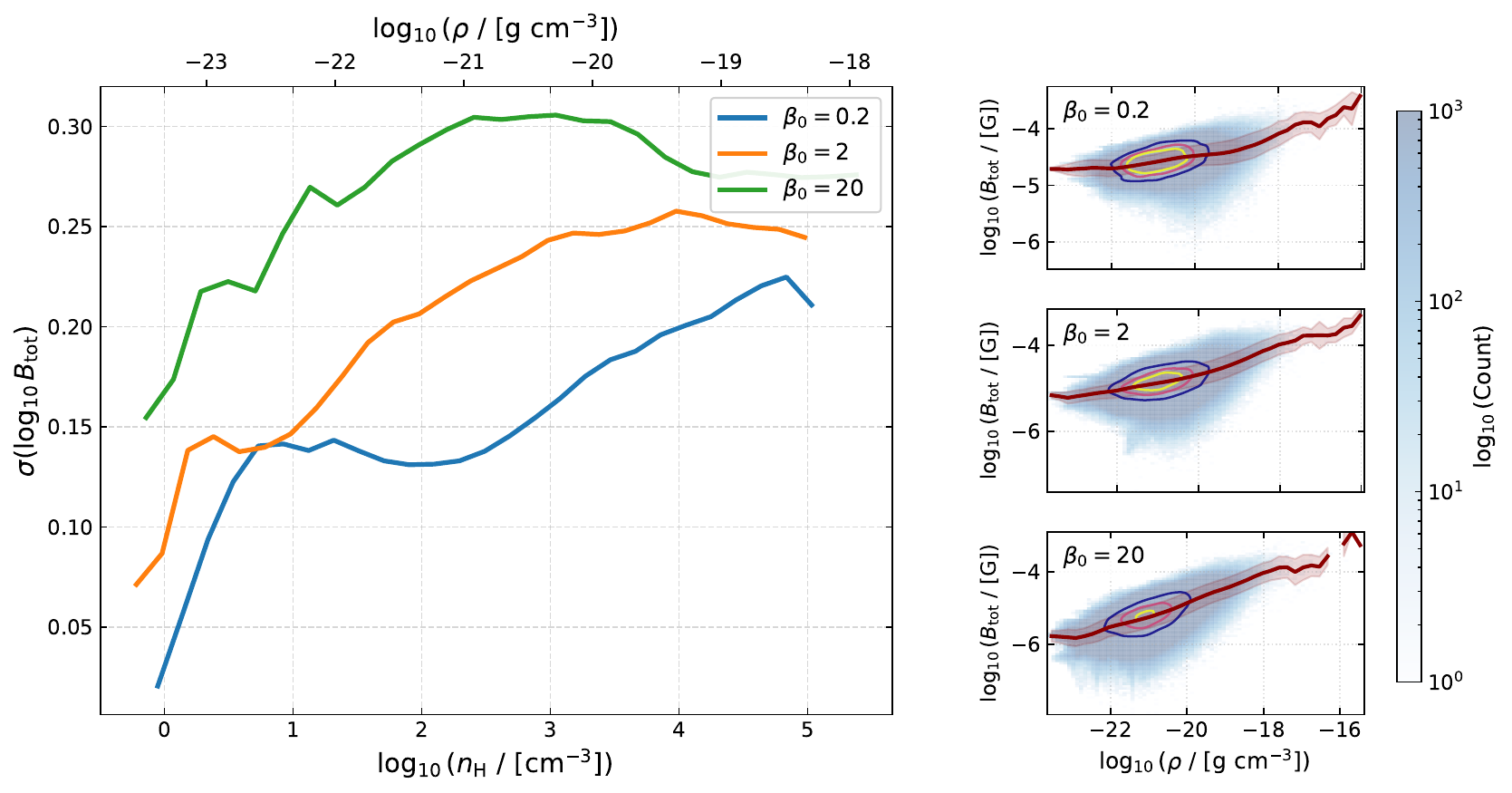}
    \caption{{\bf Intrinsic dispersion of the field at fixed density.}
    Standard deviation of $\log_{10} B_{\rm tot}$ in bins of density, measured directly in the calibrating MHD simulations \citep{2024ApJ...976..209Z} for three initial magnetisations ($\beta_0 = 0.2$, $2$, $20$).
    The dispersion increases monotonically with $\beta_0$ (weaker mean field, broader $\mathcal{M}_{\rm A}$ distribution) but shares a common shape at every magnetisation: it rises from the diffuse regime to a peak near the core-forming density $\rho \sim 10^{-19}\,\mathrm{g\,cm^{-3}}$, where the gas crosses $\mathcal{M}_{\rm A} \approx 1$ and the mix of dynamical states is broadest, and declines at higher density.
    This density structure is a prediction of the framework.
    The box matching the observed conditions ($\beta_0 = 20$) gives $\approx 0.25$ dex over the Zeeman-sampled range, a lower bound on the observed population dispersion (Section\,\ref{subsec:bic}).}
    \label{figscatter}
\end{figure}

The $B-\rho$ relation is known to carry large intrinsic scatter, both in observations \citep{2010ApJ...725..466C,2020ApJ...890..153J} and in simulations, where the field at fixed density spans more than an order of magnitude \citep{2022MNRAS.514..957S,2025MNRAS.540.2762W}. 
This scatter reflects the spread in $\mathcal{M}_{\rm A}$ at a given density rather than a failure of the mean relation, and the MHD simulation makes this quantitative. The intrinsic dispersion of $\log_{10} B_{\rm tot}$ at fixed density increases monotonically with initial magnetisation, from $\approx 0.13$ dex in the magnetically dominated box ($\beta_0 = 0.2$) to $\approx 0.25$ dex in the weakly magnetised, turbulence-dominated box ($\beta_0 = 20$), reflecting the broadening of the $\mathcal{M}_{\rm A}$ distribution as the field weakens (Fig.\,\ref{figscatter}). Using the box whose rescaling matches the observed conditions ($\beta_0 = 20$ at $0.6\,t_{\rm ff}$; \citealt{2024ApJ...976..209Z}), the intrinsic dispersion over the Zeeman-sampled range is $\approx 0.25$ dex. Line-of-sight projection ($\sim 0.2$--$0.3$ dex), measurement error, and environmental diversity add in quadrature, bringing the predicted width to $\sigma_{\rm pop} \approx 0.43$--$0.49$ dex, consistent with the fitted value (Section\,\ref{subsec:bic}). This is a consistency check on the magnitude, not an independent prediction. The \emph{shape} of the scatter is robust to this normalisation: across all three boxes ($\beta_0 = 0.2$, $2$, $20$) the dispersion rises from the diffuse end to a peak near $\rho \sim 10^{-19}\,\mathrm{g\,cm^{-3}}$ and declines at higher density (Fig.\,\ref{figscatter}), the same profile at every magnetisation. This density dependence is a prediction of the framework, distinct from the median relation: the dispersion peaks in the core-forming regime where the gas crosses $\mathcal{M}_{\rm A} \approx 1$ and the mix of dynamical states is broadest, and narrows where the gas is more uniformly magnetically (diffuse) or gravitationally (dense) dominated. This is additional information that the Gradual Transition supplies, where a single constant scatter parameter would treat the dispersion as uniform.

The residuals of our forward model are Gaussian with $\sigma \approx 0.68$ dex and no systematic trend with density (Fig.\,\ref{figD2}a); this is larger than $\sigma_{\rm pop}$ because the forward residual additionally carries the density-error and projection terms, and it describes the mean evolution while attributing the scatter to the distribution of $\mathcal{M}_{\rm A}$, consistent with the view that $B$ does not depend on $\rho$ alone \citep{2022MNRAS.514..957S}.

\section{Physically Constrained Forward Modeling}\label{ap.forward}

\subsection{Model definitions}
The Broken Power-Law (BPL) model, widely adopted in classical studies \citep{2010ApJ...725..466C,2025MNRAS.540.2762W}, assumes a discontinuous change in scaling properties at a critical density $n_0$:
\begin{equation}\label{eq.BPL}
    B(n) = 
    \begin{cases} 
    B_0 (n/n_0)^{\alpha_1} & \text{if } n_{\rm H} < n_0 \\
    B_0 (n/n_0)^{\alpha_2} & \text{if }  n_{\rm H} \ge n_0 
    \end{cases}
\end{equation}
where $\alpha_1$ and $\alpha_2$ are the slopes below and above the transition density $n_0$, and $B_0$ is the field at $n_0$.

The Gradual Transition (GT) model (Eq.\,\eqref{eq.GT}; \citealt{2024ApJ...976..209Z}; Fig.\,\ref{figD3}) has a continuous slope $k_{B-\rho}=(\rho/\rho_c)^\beta$ with $\beta=\gamma/\mathcal{K}$ set by the simulation-calibrated relations.

\begin{figure}[h]
    \centering
    \includegraphics[width=0.5\linewidth]{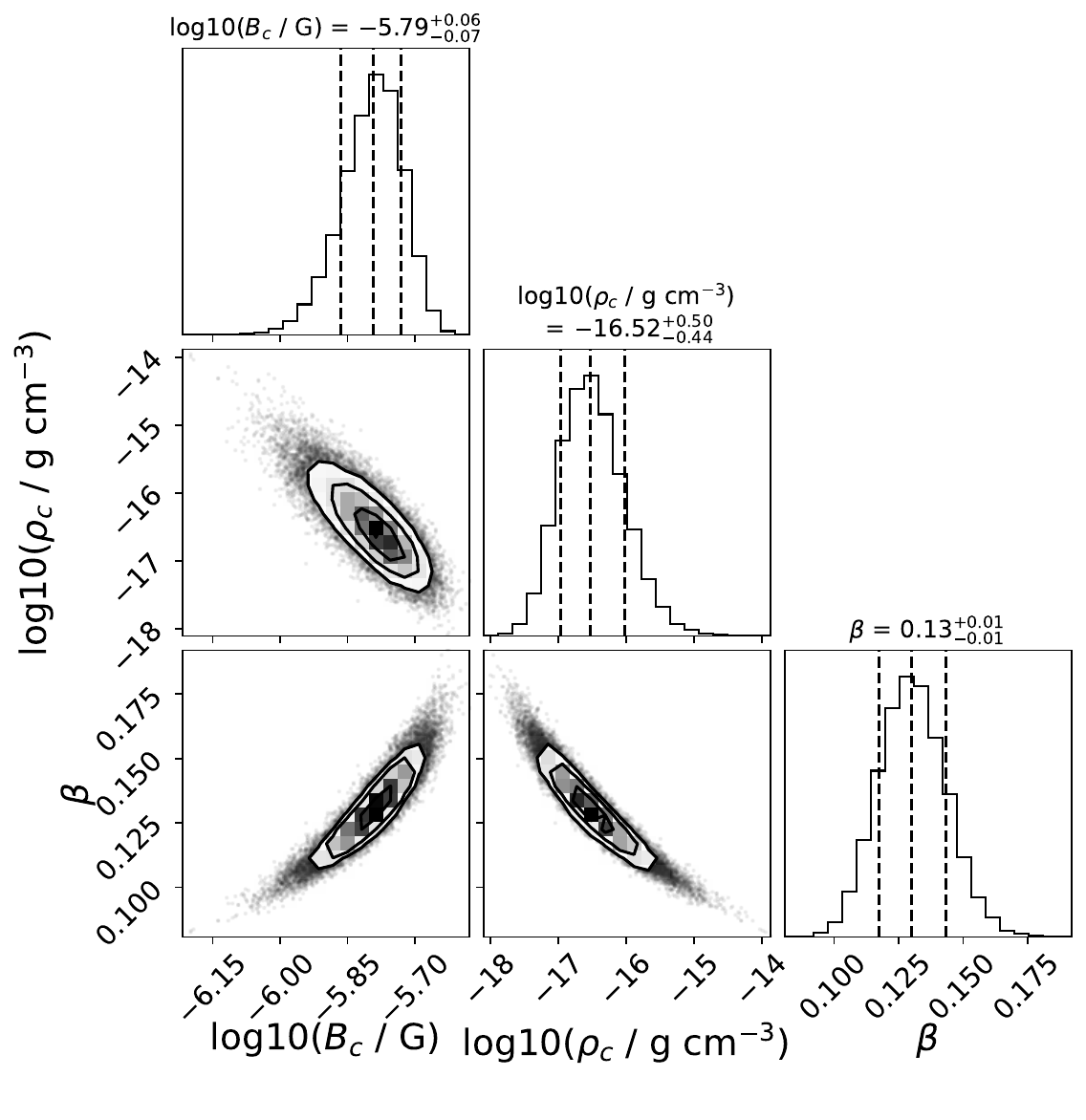}
    \caption{{\bf Posterior parameter distributions from unconstrained MCMC fitting.}
    Corner plot showing the posterior probability distributions for the three free parameters: background field strength ($\log B_c$), characteristic density ($\log \rho_c$), and scaling exponent ($\beta$). 
    Contours represent $0.5\sigma$, $1\sigma$, and $1.5\sigma$ confidence intervals.
    The resulting fit is shown by the red curves in Fig.\,\ref{figvisual}.
    }
    \label{figD4}
\end{figure}

\subsection{Forward modeling and parameter consistency check}\label{ApB2}
In the forward model $B_c$, $\rho_c$, and $\beta$ are fixed from independent constraints (below). 
To check that these fixed values are supported by the data, we separately performed an unconstrained MCMC \citep{2013PASP..125..306F} fit in which the same three parameters are left free; the corner plot of Fig.\,\ref{figD4} shows the resulting posteriors. 
For this parameter-consistency check only, we convert the observed line-of-sight field strengths to the ensemble-mean total-field scale using
$B_{\rm tot}\simeq2|B_{\rm los}|$ (Appendix\,\ref{apc8}).
$B_c$ is fixed in the forward model and free in the MCMC fit, so the posterior on $B_c$ (Fig.\,\ref{figD4}) tests whether the adopted value is supported by the data. 
The posteriors for $B_c$ and $\beta$ are statistically consistent with the independent physical values ($B_c=2.0~\mu\text{G}$, $\beta=0.147$) used in the prediction. 
The corner plot shows a correlation between $\rho_c$ and $\beta$: a larger exponent can be partly compensated by a higher transition density while preserving the high-density field strengths, which is why $\rho_c$ has the broadest posterior of the three parameters. 
The resulting covariance is propagated into the posterior predictive $B(\rho)$ relation and the reported model comparison. Although $\rho_c$ is individually broad, the predicted $B(\rho)$ relation is better constrained over the observed density range.
The full posterior values, with $1\sigma$ credible intervals, are summarised in Table\,\ref{tab.params}. 

\begin{table}[h!]
\centering
\caption{\textbf{Posterior parameters of the Gradual Transition (GT) model.} 
Values are the medians of the posterior distributions from the unconstrained MCMC fit to the unified dataset (Fig.\,\ref{figD4}), with uncertainties giving the $16$th--$84$th percentile ($1\sigma$) credible intervals. 
For reference, the physically constrained values adopted in the forward model (Fig.\,\ref{figphy}) are $\log_{10}(B_c/\text{G}) = -5.70$ ($B_c = 2.0~\mu$G), $\log_{10}(\rho_c/\text{g cm}^{-3}) \approx -16.5$, and $\beta = 0.147$, all consistent with the posteriors below.}
\label{tab.params}
\begin{tabular}{lc}
\hline
\hline
\textbf{Parameter} & \textbf{Value ($1\sigma$)} \\
\hline
$\log_{10}(B_c/\text{G})$            & $-5.79^{+0.06}_{-0.07}$ \\
$\log_{10}(\rho_c/\text{g cm}^{-3})$ & $-16.52^{+0.30}_{-0.44}$ \\
$\beta$                              & $0.13^{+0.03}_{-0.01}$ \\
\hline
\end{tabular}
\end{table}

The forward-model curve (Fig.\,\ref{figphy}a) adopts the following externally constrained parameters: 
\begin{itemize}
    \item {\bf Background Field ($B_c$)}: 
    $B_c = 2.0~\mu$G, the coherent large-scale Galactic field from pulsar rotation measures \citep{2017ARA&A..55..111H}; not fitted.
    \item {\bf Scaling Exponent ($\beta$)}: 
    From the simulation (Fig.\,\ref{figD1}), we derive the energy scaling $\mathcal{M}_{\rm A} \propto \rho^{\gamma}$ with $\gamma \approx 0.25$ from the MHD simulation \citep{2012ApJ...750...13C,2015ApJ...808...48B}.
    Combining this with the fundamental relation $\mathcal{M}_{\rm A} \propto k_{B-\rho}^{\mathcal{K}}$ ($\mathcal{K} = 1.7 \pm 0.15$) \citep{2024ApJ...976..209Z} gives the predicted exponent $\beta = \gamma / \mathcal{K} \approx 0.147$, in agreement with the MCMC posterior $\beta = 0.13^{+0.03}_{-0.01}$ (Table\,\ref{tab.params}).
    \item {\bf Characteristic Density ($\rho_c$)}: 
    We anchor $\mathcal{M}_{\rm A}=1$ at $n_{\rm trans} \approx 1800$ cm$^{-3}$ \citep{2025ApJ...988..132Y}; combined with $\beta=0.147$ this fixes $\rho_c \approx 10^{-16.5}$\,g\,cm$^{-3}$.
\end{itemize}

\begin{table}[h!]
\centering
\caption{\textbf{Specification-controlled Bayesian model comparison.}
$\Delta \text{BIC} = \text{BIC}_{\text{BPL}} - \text{BIC}_{\text{GT}}$ between the generalised four-parameter Broken Power-Law and the three-parameter Gradual Transition model; positive values favour GT, and $|\Delta\text{BIC}| > 10$ is conventionally decisive \citep{1995JASA...90..773K}.
All rows use the symmetric lognormal-median likelihood of Appendix\,\ref{Ap.BIC}; identical nuisance parameters are attached to both models in every row, so the BIC penalty difference is $\ln N$ throughout and only the likelihoods respond to the specification.
$R$ is the density-uncertainty factor ($\sigma_{\log\rho} \approx \log R/2$): $R = 2$ \citep{2010ApJ...725..466C} and $R = 9.3$ \citep{2020ApJ...890..153J}.
$f_p$ is the pulsar-branch suppression factor (projection, path averaging, field reversals).
No specification reaches a decisive preference ($|\Delta\text{BIC}|>10$) for either functional form; the sign of $\Delta\text{BIC}$ itself depends on the treatment of the projection factor $f_p$, from $+4.3$ (favouring GT, $f_p$ fixed) to $-8.5$ (favouring BPL, $f_p$ prior), which we take as direct evidence that the dense data do not determine the functional form.}
\label{tab:bic_comparison}
\begin{tabular}{lcc}
\hline
\hline
\textbf{Specification} & \textbf{$\Delta$BIC ($R=2$)} & \textbf{$\Delta$BIC ($R=9.3$)} \\
\hline
Zeeman-only ($N=137$), symmetric likelihood & $-1.1$ & $-4.9$ \\
\hline
Unified ($N=761$), $f_p$ free & $-5.4$ & $-6.9$ \\
Unified ($N=761$), $f_p$ prior $\mathcal{N}(\log_{10} 0.4,\,0.15\,\mathrm{dex})$ & $-8.1$ & $-8.5$ \\
Unified ($N=761$), $f_p = 0.5$ fixed & $+1.8$ & $+4.3$ \\
Unified ($N=761$), $f_p$ prior, two high-leverage sources removed & $-2.3$ & $-1.9$ \\
\hline
\end{tabular}
\end{table}

\begin{table}[h!]
\centering
\caption{\textbf{A-priori predictions versus unconstrained fits.}
Each GT parameter is fixed by external physics \emph{before} any fit to the present data (left column); the right column gives the value recovered by the unconstrained unified fit (Section\,\ref{subsec:bic}, range over $R$ and over the $f_p$ treatments of Table\,\ref{tab:bic_comparison}).
The Gradual Transition is the only model whose parameters are set in advance by external relations rather than calibrated to the data.}
\label{tab:predictions}
\begin{tabular}{lcc}
\hline
\hline
\textbf{Parameter} & \textbf{Predicted (a priori)} & \textbf{Fitted (unified)} \\
\hline
Scaling exponent $\beta$ & $0.147$ \citep{2024ApJ...976..209Z} & $0.15$--$0.21$ \\
Background floor $B_c$ & $2.0\,\mu$G  & $1.9$--$2.3\,\mu$G \\
Suppression factor $f_p$ & $0.3$--$0.6$ (projection) & $0.54$--$0.61$ \\
\hline
\end{tabular}
\end{table}

\subsection{Dense-gas verdict under two likelihoods}\label{ap.densegas}
Under the flat-envelope likelihood of \citet{2010ApJ...725..466C} the Zeeman-only comparison ($N=137$) appears to favour the broken form ($\Delta\text{BIC}=-8.9$ at $R=2$, $-15.1$ at $R=9.3$). An injection--recovery calibration shows this to be misleading: generating mock catalogues from each best-fit model and refitting both, the observed statistic lies outside the distribution produced by \emph{either} model, so under this likelihood the comparison is dominated by misspecification rather than by the shape of the relation. A per-point decomposition localises the apparent preference to a small set of high-leverage structures the flat envelope cannot generate (a cluster of OH non-detections sharing one assigned density, and two measurements lying $\gtrsim8\sigma$ above both fitted envelopes), and to the empty sampling interval $(280,480)$\,cm$^{-3}$ in which the fitted break density falls. Under the symmetric lognormal-median likelihood (Section\,\ref{ap.likespec}) the preference collapses to $\Delta\text{BIC}=-1.1$ ($R=2$): the two median curves coincide within the population scatter wherever the data sample them.

\section{Statistical Analysis and Robustness}\label{Ap.BIC}

This appendix documents the diagnostic chain behind the specification-controlled comparison of Section\,\ref{subsec:bic}.
The Bayesian Information Criterion is $\text{BIC} = k \ln(N) - 2 \ln(\hat{L})$ \citep{1978AnSta...6..461S}.
For the GT model the physical parameters are ($B_c$, $\rho_c$, $\beta$), $k = 3$; for the generalised BPL they are ($B_0$, $n_0$, $\alpha_1$, $\alpha_2$), $k = 4$; each specification below attaches an identical set of nuisance parameters to both models, so the BIC penalty difference is $\ln N$ in every comparison and only the maximised likelihoods respond to the specification.

\subsection{Likelihood specifications}\label{ap.likespec}
We consider two families.
(i) The flat-envelope convolution of \citet{2010ApJ...725..466C} and \citet{2020ApJ...890..153J}: the model curve is interpreted as the upper envelope $B_{\rm max}(n)$, the total field is drawn flat on $[0, B_{\rm max}]$, the line-of-sight component follows isotropic projection ($B_z = B\cos\theta$), Gaussian measurement noise is convolved analytically, and the density uncertainty is marginalised with a log-uniform factor-of-$R$ prior.
(ii) A symmetric population likelihood: both laws are treated as the \emph{median} of a lognormal $B_{\rm tot}$ distribution with population dispersion $\sigma_{\rm pop}$, projected by $\cos\theta$ for single-cloud Zeeman measurements, with the same noise convolution and density marginalisation.
For the pulsar branch the path-averaged $\langle B_\parallel \rangle$ is Gaussianised by the central limit theorem, so the single-cloud projection is inapplicable; we use a log-Gaussian branch with a global suppression factor $f_p$ (projection, path averaging, field reversals) and total scatter $\sigma_p$, the density error entering through the local model slope.
To remove a ridge degeneracy between $B_c$ and $\beta$ on windowed data, the GT model is fitted in a pivot parameterisation, $B_p \equiv B(n_{\rm H} = 10^3\,{\rm cm^{-3}})$, with $B_c$ recovered afterwards.
The effective-variance Gaussian likelihood on $2|B_{\rm los}|$ used for model selection in the original submission is superseded by these specifications; its original $\Delta\text{BIC}$ values have therefore been withdrawn. The factor-of-two ensemble conversion is retained only for the separate parameter-consistency and blind-extrapolation fits described in Appendices~\ref{ApB2} and~\ref{apc8}.

\subsection{Injection--recovery calibration of the envelope comparison}\label{ap.inject}
Under specification (i) the Zeeman-only comparison returns $\Delta\text{BIC} = -8.9$ ($R = 2$) and $-15.1$ ($R = 9.3$), apparently favouring the broken form.
To calibrate this statistic we generated mock datasets from each best-fit model (flat $P(B)$, isotropic projection, catalogue noise, and the catalogue density sampling) and refitted both models under the same procedure.
With GT as the truth, the recovered $\Delta\text{BIC}$ has median $+3.5$ ($R = 2$; 16--84\% range $+1.3$ to $+6.1$) and none of twenty mocks reaches the observed value; with BPL as the truth the median is $-0.1$ and two of twenty reach it.
At $R = 9.3$ the observed $-15.1$ lies outside all forty mocks of both truths.
The observed statistic is therefore not typical of either generative model: under specification (i) the comparison is dominated by likelihood misspecification rather than by the shape of the $B$--$\rho$ relation.

\subsection{Per-point decomposition under the envelope likelihood}\label{ap.perpoint}
Decomposing the likelihood gap point by point localises it in three structures:
(a) a cluster of OH non-detections sharing a single assigned survey density ($n_{\rm H} \approx 2\times10^3\,{\rm cm^{-3}}$), contributing $\approx -2.3$ of the total $-6.9$ log-likelihood units, whose correlated density assignment the framework treats as independent errors;
(b) two measurements lying $\gtrsim 8\sigma$ above \emph{both} fitted envelopes ($n_{\rm H} = 5\times10^3\,{\rm cm^{-3}}$, $|B_z| = 480\,\mu$G; $n_{\rm H} = 6.3\times10^4\,{\rm cm^{-3}}$, $|B_z| = 500\,\mu$G), events of essentially zero probability under a flat $P(B)$, contributing of order $\pm 1$ unit each, with signs set merely by which envelope lies marginally closer;
(c) two HI detections in the middle of that tracer window.
Excluding the HI window leaves $\Delta\text{BIC} = -5.0$; excluding the two envelope-violating points leaves $-7.2$.
These are precisely the data features that the flat-envelope generative model cannot produce, consistent with the injection result above.
We further note that the fitted BPL break density, $n_0 \approx 380\,{\rm cm^{-3}}$ at $R = 2$, falls inside the empty sampling interval $(280, 480)\,{\rm cm^{-3}}$ between the HI and OH windows, where no datum can penalise the kink.

\subsection{Symmetric likelihood: collapse of the dense-gas verdict and unified decomposition}\label{ap.symmetric}
On the unified dataset we treat the pulsar suppression factor $f_p$ (projection, path averaging, field reversals) in three ways: free, constrained by the physical prior $\log_{10}f_p\sim\mathcal{N}(\log_{10}0.4,\,0.15)$, and fixed to the isotropic-median value $0.5$ (Table\,\ref{tab:bic_comparison}).
Under specification (ii) the Zeeman-only preference collapses ($\Delta\text{BIC} = -1.1$ at $R = 2$; $-4.9$ at $R = 9.3$); the two fitted median curves coincide within the population scatter over $n_{\rm H} \approx 10$--$10^4\,{\rm cm^{-3}}$, and the best-fit BPL itself softens ($\alpha_1 = 0.19$, with the break migrating inside the OH data rather than hiding in the sampling gap).
On the unified dataset the residual preference (Table\,\ref{tab:bic_comparison}) decomposes as follows ($R = 2$, $f_p$ prior): $\approx 80\%$ of the likelihood gap is carried by the dense-gas branch, the remainder by the pulsar branch and the prior term.
The shared-density cluster of OH non-detections is exonerated here: it mildly favours the unified GT fit.
The cross-window consistency cost, the additional dense-gas penalty incurred when the pulsars anchor the GT floor at its physical value, is carried almost entirely by the two legacy high-leverage sources: removing them reduces it from $2.9$ to $0.3$ log-likelihood units, i.e.\ the smooth law spans the ten decades of density without internal strain apart from these two measurements.
Removing these two sources and refitting under the physical-prior treatment reduces the broken-form preference from $\Delta\text{BIC}\approx-8$ to $-2.3$ ($R=2$) and to $-1.9$ ($R=9.3$, inconclusive), and the fitted GT exponent relaxes from $\beta\approx0.19$ towards its predicted value ($\beta=0.165$--$0.176$ against the a-priori $0.147$); the cross-window strain was carried by the two measurements, not by the smooth law.
A complementary control fixes the projection factor to $f_p=0.5$ for both models. The broken power law can then no longer absorb the pulsar--Zeeman level offset through $f_p$ and is forced to a non-zero diffuse slope, $\alpha_1\approx0.06$--$0.08$ (against $\alpha_1=0$ when $f_p$ is free), in the direction of the rising diffuse segment found independently by \citet{2026arXiv260306838W}; its second slope parameter then earns no likelihood improvement over the single-exponent GT form, and the parsimony term turns the comparison to $\Delta\text{BIC}=+1.8$ ($R=2$) and $+4.3$ ($R=9.3$).
The remaining preference is dominated by strong-field detections at $n_{\rm H} \approx (1\text{--}7)\times10^{5}\,{\rm cm^{-3}}$, whose levels prefer a locally constant slope $k_{B-\rho} \approx 0.65$ over a running slope; the physical reading, saturation of the slope evolution in the trans-Alfv\'enic regime, is discussed in Sections\,\ref{subsec:bic} and \ref{subsec:regimes}.
Across all treatments the fitted characteristic density sits about a decade below its a-priori value ($\rho_c\sim10^{-17.5}$ against $10^{-16.5}$), the data pulling $\rho_c$ downward through the same strong-field group; for this reason we treat $\rho_c$, unlike $\beta$ and $B_c$, as physically anchored rather than as a recovered prediction.

\subsection{Pulsar internal-slope caveat}\label{ap.dmshear}
The pulsar density and field estimates share the dispersion measure ($n \propto {\rm DM}/d$, $B_\parallel \propto {\rm RM}/{\rm DM}$), so DM errors are anti-correlated between the two axes and shear the pulsar cloud toward slope $-1$.
The internal ordinary-least-squares slope of the pulsar cloud is $-0.065$, negative where the BPL predicts $0$ and the GT predicts $\approx +0.05$.
The pulsar-branch contribution to $\Delta\text{BIC}$ should therefore be read as an upper limit on the evidence against slope running in the diffuse regime.

\begin{figure}[h]
    \centering
    \includegraphics[width=0.46\linewidth]{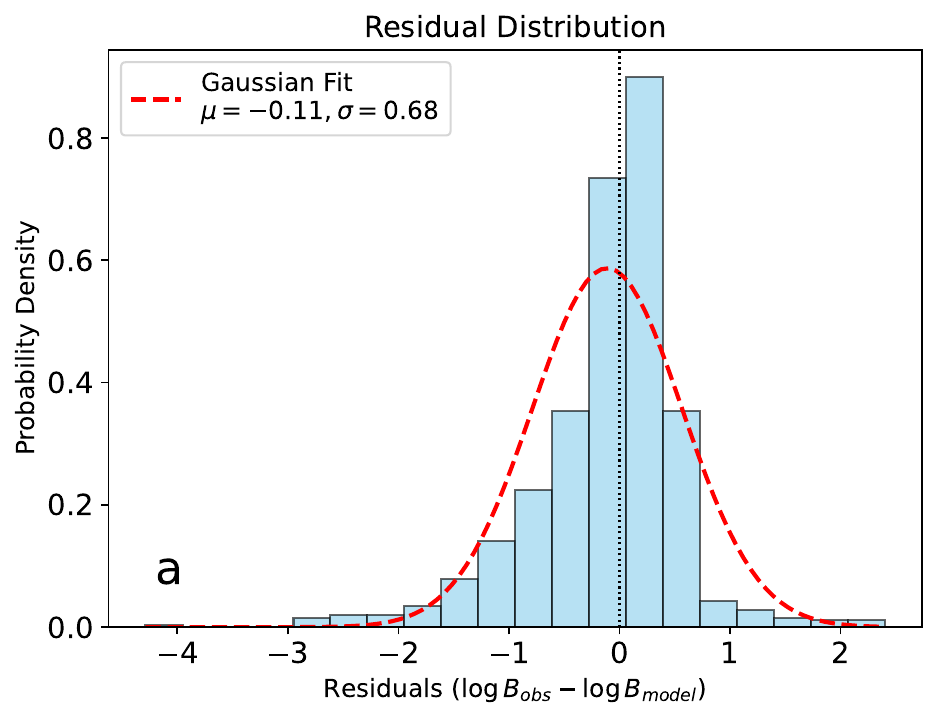}
    \includegraphics[width=0.5\linewidth]{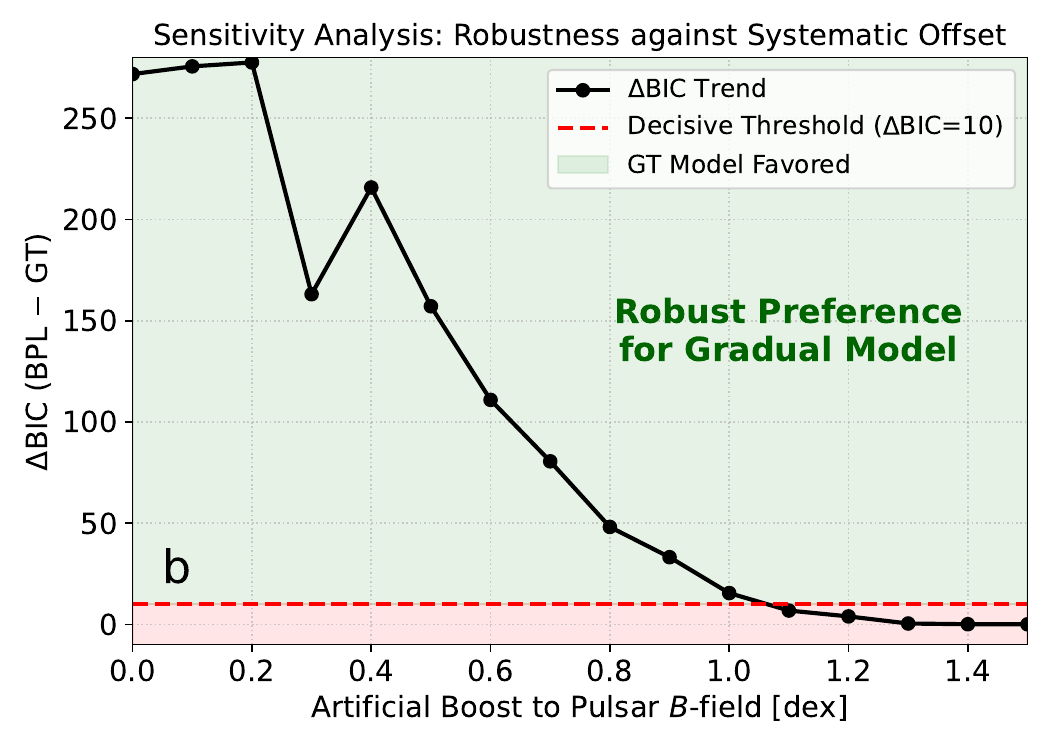}
    \caption{{\bf Statistical diagnostics and robustness of Gradual Transition model.}
    a, Histogram of the residuals ($\Delta \log B = \log B_{\text{obs}} - \log B_{\text{model}}$) derived from the forward modeling, complementing Fig.\,\ref{figphy}b. 
    The distribution is well-described by a Gaussian function (red dashed line) centered near zero ($\mu \approx -0.11$) with a standard deviation of $\sigma \approx 0.68$ dex, consistent with the intrinsic scatter of the $B-\rho$ relation.
    b, Stress test of cross-tracer normalisation: $\Delta \text{BIC}$ (effective-variance framework, $R = 9.3$) as pulsar fields are boosted by $10^x$. The nuisance parameter $f_p$ in Table\,\ref{tab:bic_comparison} supersedes this test.
    }
    \label{figD2}
\end{figure}

\subsection{Cross-tracer normalisation}
Potential systematic offsets in the pulsar measurements (field reversals, ionisation bias, projection) are handled by the explicit suppression parameter $f_p$, reported under three treatments in Table\,\ref{tab:bic_comparison}: fully free (most conservative, in which the level continuity between the pulsar and Zeeman populations carries no evidential weight), a physical prior $\log_{10} f_p \sim \mathcal{N}(\log_{10} 0.4,\,0.15\,\text{dex})$ encoding the expected projection and path-averaging suppression, and the fixed isotropic-median convention $f_p = 0.5$.
The earlier boost-style stress test (Fig.\,\ref{figD2}b), computed in the original effective-variance framework, is retained for reference.

\subsection{Blind extrapolation}\label{ap.blind}
The blind test summarised in Section\,\ref{subsec:blind} fits the GT model exclusively to the dense-gas Zeeman sample ($n_{\rm H} \gtrsim 100$ cm$^{-3}$), with the diffuse pulsar measurements excluded. 
The best-fit curve is then extrapolated, without any further adjustment, across the four decades of density that separate the dense gas from the diffuse ionised medium. 
The result is shown in Fig.\,\ref{figE2}: the curve calibrated on dense cores passes through the pulsar measurements. 
The GT framework \citep{2024ApJ...976..209Z} predates the comprehensive pulsar compilation \citep{2025MNRAS.tmp..512S}, so the extrapolation is a prediction of data the model has not seen.

\subsection{Treatment of geometric projection}\label{apc8}

Zeeman and Faraday-rotation measurements probe the line-of-sight magnetic-field component, $B_{\rm los}$, whereas the Gradual Transition relation is formulated in terms of the underlying magnetic-field strength. For direct presentation of the observational compilation, we retain the measured line-of-sight field strengths, $|B_{\rm los}|$, so that the pulsar and Zeeman data are shown on the same observational footing.

Projection is treated differently according to the purpose of the analysis. In the Bayesian model comparison of Section\,\ref{subsec:bic}, no fixed conversion from $B_{\rm los}$ to $B_{\rm tot}$ is applied. For the Zeeman sample, the symmetric likelihood treats $B_{\rm tot}$ as the latent field strength and explicitly marginalises over the orientation of individual fields. For the pulsar sample, whose RM/DM estimates are path-averaged, projection, field reversals, and path averaging are absorbed into the nuisance factor $f_p$, reported under three treatments in Table\,\ref{tab:bic_comparison}.

For the separate parameter-consistency MCMC of Appendix\,\ref{ApB2}, and for the blind-extrapolation fits shown in Fig.\,\ref{figE2}, we instead place the measurements on an ensemble-mean total-field scale. For a large ensemble of randomly oriented magnetic fields, the statistical expectation is $\langle B_{\rm tot}\rangle \simeq 2\langle |B_{\rm los}|\rangle$ \citep{1999ApJ...520..706C}. We therefore use $B_{\rm tot}\simeq2|B_{\rm los}|$ in these fits, allowing the recovered parameters $(B_c,\rho_c,\beta)$ and the extrapolated relation to be compared directly with the physically constrained Gradual Transition model. This factor-of-two conversion is not used in the Bayesian model comparison.

\begin{figure}[h]
    \centering
    \includegraphics[width=0.5\linewidth]{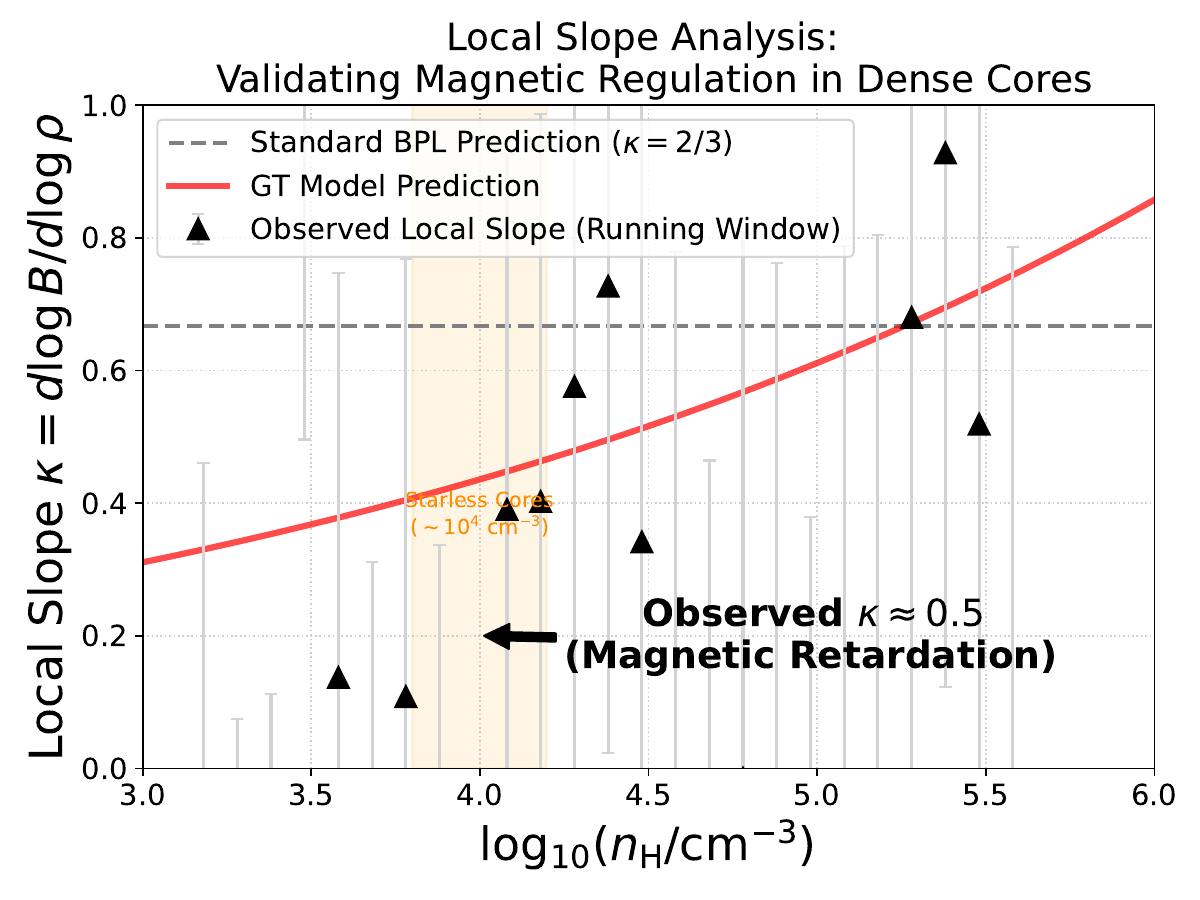}
    \caption{{\bf Local slope of the $B$--$\rho$ relation in the high-density regime.}
    Running-window estimate of $k_{B-\rho} = d\log B / d\log n_{\rm H}$ (width 0.6 dex). Triangles: local mean slope; gray bars: $1\sigma$ scatter. The BPL predicts a jump to $k_{B-\rho}=2/3$ (gray dashed) at $n_{\rm H} > 300$ cm$^{-3}$; the observed slopes in the starless-core regime ($n_{\rm H} \sim 10^{4}$ cm$^{-3}$, yellow shading) lie below this value and are consistent with the GT curve (red). }
    \label{figE1}
\end{figure}

\subsection{$B$-$\rho$ slope analysis}\label{ap.local}
As a model-independent complement to the global comparison of Section\,\ref{subsec:bic}, we measured the $B$-$\rho$ slope $k_{B-\rho} = d\log B/d\log n_{\rm H}$ directly from the data, asking whether it reaches the isotropic-collapse value of $2/3$ \citep{1966MNRAS.133..265M} or remains shallower \citep{1999ASIC..540..305M} at starless-core densities ($n_{\rm H} \sim 10^4$ cm$^{-3}$; \citealt{2007ARA&A..45..339B}; Fig.\,\ref{figE1}).
The dataset was sorted by density and $k_{B-\rho}$ calculated in a sliding window of width 0.6 dex with a step of 0.2 dex.
Projection ($B_{\rm los} = B_{\rm tot}\sin\theta$) adds scatter but, for orientations uncorrelated with density, does not bias the ensemble-mean slope.

Numerical differentiation amplifies noise, so the per-window uncertainties are large (Fig.\,\ref{figE1}), but in the range $10^{3.8} < n_{\rm H} < 10^{4.2}$ cm$^{-3}$, the derived slopes cluster around $k_{B-\rho} \approx 0.4-0.5$, below the BPL value of $k_{B-\rho}=2/3$ \citep{1966MNRAS.133..265M} and consistent with the Gradual Transition. We caution that the Zeeman compilation at $n_{\rm H} \gtrsim 10^4$ cm$^{-3}$ is incomplete and biased towards detections of stronger fields \citep{2010ApJ...725..466C, 2020ApJ...890..153J}; the running-window estimate inherits this selection effect and is therefore a consistency check rather than a stand-alone proof. What it does establish, independently of any global model, is that a one-step jump from $k_{B-\rho}=0$ to $2/3$ is not what the available data support. This is the $B$-$\rho$ slope of the Gradual Transition at starless-core densities, not a single global exponent, and is in good agreement with the $k_{B-\rho}\approx1/2$ obtained by \citet{2015MNRAS.451.4384T} on refitting the Crutcher data with relaxed density uncertainties: both place the high-density slope below the isotropic value of $2/3$.

\section{Magnetic Energy Spectrum: Derivation}\label{ap.spectrum}

This appendix gives the technical derivation of the magnetic energy spectrum $E_B(k)$ discussed in Section\,\ref{sec:prediction}. 
We map the spatial scale $r$ to wavenumber as $k = 1/r$ and connect $r$ to density through the size–density relation $\rho \propto r^{-1.1}$ of the molecular ISM \citep{1981MNRAS.194..809L}, normalised to $n_{\rm H}\approx 10^3$ cm$^{-3}$ at $r=1$ pc \citep{1987ApJ...319..730S, 2000prpl.conf...97W}. 
The molecular-cloud size--density relation $\rho\propto r^{-1.1}$ \citep{1981MNRAS.194..809L} implies $M(r)\propto\rho\,r^3\propto r^{1.9}$. Identifying the mass--size exponent with the effective fractal dimension therefore gives $D_f\simeq1.9$, and the spatial intermittency of dense structures is encoded in a volume filling factor $f_v(k)\propto k^{D_f-3}\simeq k^{-1.1}$.
The magnetic energy spectrum then reads
\begin{equation}\label{eq.spectrum}
    E_B(k) \propto B(\rho)^2\,k^{-1}\,f_v(k),
\end{equation}
with $B(\rho)$ given by the Gradual Transition model, Eq.\,\eqref{eq.GT}. 
Thus, the scale mapping and filling-factor scaling are both fixed by the adopted Larson relation rather than fitted to the present magnetic-field data. 
To propagate the parameter uncertainty, we vary the characteristic density $\rho_c$ across its MCMC posterior (Section\,\ref{subsec:posterior}; Table\,\ref{tab.params}) and plot a family of curves in Fig.\,\ref{figD5}. 
The physical content of the resulting prediction is discussed in Section\,\ref{sec:prediction}.

\end{document}